\documentclass[12pt]{article}
\usepackage{epsfig}
\usepackage{amsmath}
\usepackage{hhline}
\usepackage{amssymb}
\usepackage{times}
\usepackage{cite}

\newlength{\dinwidth}
\newlength{\dinmargin}
\begin{document}
% The rest
\newcommand{\pom}{{I\!\!P}}
\newcommand{\reg}{{I\!\!R}}
\newcommand{\slowpi}{\pi_{\mathit{slow}}}
\newcommand{\fiidiii}{F_2^{D(3)}}
\newcommand{\fiidiiiarg}{\fiidiii\,(\beta,\,Q^2,\,x)}
\newcommand{\n}{1.19\pm 0.06 (stat.) \pm0.07 (syst.)}
\newcommand{\nz}{1.30\pm 0.08 (stat.)^{+0.08}_{-0.14} (syst.)}
\newcommand{\fiidiiiful}{F_2^{D(4)}\,(\beta,\,Q^2,\,x,\,t)}
\newcommand{\fiipom}{\tilde F_2^D}
\newcommand{\ALPHA}{1.10\pm0.03 (stat.) \pm0.04 (syst.)}
\newcommand{\ALPHAZ}{1.15\pm0.04 (stat.)^{+0.04}_{-0.07} (syst.)}
\newcommand{\fiipomarg}{\fiipom\,(\beta,\,Q^2)}
\newcommand{\pomflux}{f_{\pom / p}}
\newcommand{\nxpom}{1.19\pm 0.06 (stat.) \pm0.07 (syst.)}
\newcommand {\gapprox}
   {\raisebox{-0.7ex}{$\stackrel {\textstyle>}{\sim}$}}
\newcommand {\lapprox}
   {\raisebox{-0.7ex}{$\stackrel {\textstyle<}{\sim}$}}
\def\gsim{\,\lower.25ex\hbox{$\scriptstyle\sim$}\kern-1.30ex%
\raise 0.55ex\hbox{$\scriptstyle >$}\,}
\def\lsim{\,\lower.25ex\hbox{$\scriptstyle\sim$}\kern-1.30ex%
\raise 0.55ex\hbox{$\scriptstyle <$}\,}
\newcommand{\pomfluxarg}{f_{\pom / p}\,(x_\pom)}
\newcommand{\dsf}{\mbox{$F_2^{D(3)}$}}
\newcommand{\dsfva}{\mbox{$F_2^{D(3)}(\beta,Q^2,x_{I\!\!P})$}}
\newcommand{\dsfvb}{\mbox{$F_2^{D(3)}(\beta,Q^2,x)$}}
\newcommand{\dsfpom}{$F_2^{I\!\!P}$}
\newcommand{\gap}{\stackrel{>}{\sim}}
\newcommand{\lap}{\stackrel{<}{\sim}}
\newcommand{\fem}{$F_2^{em}$}
\newcommand{\tsnmp}{$\tilde{\sigma}_{NC}(e^{\mp})$}
\newcommand{\tsnm}{$\tilde{\sigma}_{NC}(e^-)$}
\newcommand{\tsnp}{$\tilde{\sigma}_{NC}(e^+)$}
\newcommand{\st}{$\star$}
\newcommand{\sst}{$\star \star$}
\newcommand{\ssst}{$\star \star \star$}
\newcommand{\sssst}{$\star \star \star \star$}
\newcommand{\tw}{\theta_W}
\newcommand{\sw}{\sin{\theta_W}}
\newcommand{\cw}{\cos{\theta_W}}
\newcommand{\sww}{\sin^2{\theta_W}}
\newcommand{\cww}{\cos^2{\theta_W}}
\newcommand{\trm}{m_{\perp}}
\newcommand{\trp}{p_{\perp}}
\newcommand{\trmm}{m_{\perp}^2}
\newcommand{\trpp}{p_{\perp}^2}
\newcommand{\alp}{\alpha_s}

\newcommand{\alps}{\alpha_s}
\newcommand{\sqrts}{$\sqrt{s}$}
\newcommand{\LO}{$O(\alpha_s^0)$}
\newcommand{\Oa}{$O(\alpha_s)$}
\newcommand{\Oaa}{$O(\alpha_s^2)$}
\newcommand{\PT}{p_{\perp}}
\newcommand{\JPSI}{J/\psi}
\newcommand{\sh}{\hat{s}}
\newcommand{\uh}{\hat{u}}
\newcommand{\MP}{m_{J/\psi}}
\newcommand{\PO}{I\!\!P}
\newcommand{\xbj}{x}
\newcommand{\xpom}{x_{\PO}}
\newcommand{\ttbs}{\char'134}
\newcommand{\xpomlo}{3\times10^{-4}}
\newcommand{\xpomup}{0.05}
\newcommand{\dgr}{^\circ}
\newcommand{\pbarnt}{\,\mbox{{\rm pb$^{-1}$}}}
\newcommand{\gev}{\,\mbox{GeV}}
\newcommand{\WBoson}{\mbox{$W$}}
\newcommand{\fbarn}{\,\mbox{{\rm fb}}}
\newcommand{\fbarnt}{\,\mbox{{\rm fb$^{-1}$}}}
%
% Some useful tex commands
%
\newcommand{\qsq}{\ensuremath{Q^2} }
\newcommand{\gevsq}{\ensuremath{\mathrm{GeV}^2} }
\newcommand{\et}{\ensuremath{E_t^*} }
\newcommand{\rap}{\ensuremath{\eta^*} }
\newcommand{\gp}{\ensuremath{\gamma^*}p }
\newcommand{\dd}{\mathrm{d}}

\newcommand{\dsiget}{\ensuremath{{\rm d}\sigma_{ep}/{\rm d}E_t^*} }
\newcommand{\dsigrap}{\ensuremath{{\rm d}\sigma_{ep}/{\rm d}\eta^*} }
% Journal macro
\def\Journal#1#2#3#4{{#1} {\bf #2} (#3) #4}
\def\NCA{\em Nuovo Cimento}
\def\NIM{\em Nucl. Instrum. Methods}
\def\NIMA{{\em Nucl. Instrum. Methods} {\bf A}}
\def\NPB{{\em Nucl. Phys.}   {\bf B}}
\def\PLB{{\em Phys. Lett.}   {\bf B}}
\def\PRL{\em Phys. Rev. Lett.}
\def\PRD{{\em Phys. Rev.}    {\bf D}}
\def\ZPC{{\em Z. Phys.}      {\bf C}}
\def\EJC{{\em Eur. Phys. J.} {\bf C}}
\def\CPC{\em Comp. Phys. Commun.}

\begin{titlepage}

\noindent
%Date: 18/12/2001             \\
%Version: 4.0 \\
%Editors: Sascha Caron, Christoph Berger           \\
%Referees: Daniel Pitzl, Stefan Schmitt    \\

\begin{flushleft}
  {\tt DESY 01-225} \hfill {\tt ISSN 0418-9833} \\
  {\tt December 2001}  
\end{flushleft}

\vspace*{3cm}

\begin{center}
\begin{Large}

{\bf Measurement of Dijet Cross Sections in Photoproduction\\
 at HERA}

\vspace{2cm}

H1 Collaboration 

\end{Large}
\end{center}

\vspace{2cm}

\begin{abstract}
\noindent
Dijet cross sections as functions of
several jet observables are measured in
photoproduction using the H1 detector at HERA.
The data sample
comprises $e^+p$ data with an integrated luminosity of 34.9 pb$^{-1}$.
Jets
are selected using the inclusive $k_{\bot}$ algorithm
with a minimum transverse energy of $25~\gev$ for the leading jet.
The phase space covers longitudinal proton momentum fraction
$x_p$ and photon longitudinal momentum fraction $x_{\gamma}$ 
in the ranges $0.05<x_p<0.6$ and $0.1<x_{\gamma}<1$. 
The predictions of next-to-leading order perturbative QCD, including
recent photon and proton parton densities, are found to be compatible
with the data in a wide kinematical range.

\end{abstract}
\vspace{1.5cm}

\begin{center}
To be submitted to Eur. Phys. J. C
\end{center}

\end{titlepage}

%

%          COPY THE AUTHOR AND INSTITUTE LISTS

%       AT THE TIME OF THE T0-TALK INTO YOUR AREA

%

% from /h1/iww/ipublications/h1auts.tex

\begin{flushleft}
  %-- H1AUTS Author list by names 
%-- Status: Thu Oct  4 18:26:02 MET DST 2001  Number of authors = 331 

C.~Adloff$^{33}$,              %WUPP-LEFT      07/01           Adloff              
V.~Andreev$^{24}$,             %LPI -PD        8/88            Andreev             
B.~Andrieu$^{27}$,             %ECPL-LEFT      09/01           Andrieu             
T.~Anthonis$^{4}$,             %ANTW-ST        11/99           Anthonis            
V.~Arkadov$^{35}$,             %ZEUT-LEFT      10/0            Arkadov             
A.~Astvatsatourov$^{35}$,      %ZEUT-ST        02/98           Astvatsatourov      
A.~Babaev$^{23}$,              %ITEP-PD        8/88            Babaev              
J.~B\"ahr$^{35}$,              %ZEUT-PD        8/88            Baehr               
P.~Baranov$^{24}$,             %LPI -PD        8/88            Baranovp            
E.~Barrelet$^{28}$
% ...

%-- H1 Institutes 
\bigskip{\it
 $ ^{1}$ I. Physikalisches Institut der RWTH, Aachen, Germany$^{ a}$ \\
% ...
$ ^{4}$ Inter-University Institute for High Energies ULB-VUB, Brussels;
          Universiteit Antwerpen (UIA), Antwerpen; Belgium$^{ c}$ \\
% ...
$ ^{23}$ Institute for Theoretical and Experimental Physics,
          Moscow, Russia$^{ l}$ \\
 $ ^{24}$ Lebedev Physical Institute, Moscow, Russia$^{ e}$ \\
% ...
$ ^{27}$ LPNHE, Ecole Polytechnique, IN2P3-CNRS, Palaiseau, France \\
 $ ^{28}$ LPNHE, Universit\'{e}s Paris VI and VII, IN2P3-CNRS,
          Paris, France \\
% ...
$ ^{33}$ Fachbereich Physik, Bergische Universit\"at Gesamthochschule
          Wuppertal, Wuppertal, Germany \\
% ...
$ ^{35}$ DESY, Zeuthen, Germany \\
% ...
$ ^a$ Supported by the Bundesministerium f\"ur Bildung und Forschung, FRG,
      under contract numbers 05 H1 1GUA /1, 05 H1 1PAA /1, 05 H1 1PAB /9,
      05 H1 1PEA /6, 05 H1 1VHA /7 and 05 H1 1VHB /5 \\
% ...
$ ^c$ Supported by FNRS-FWO-Vlaanderen, IISN-IIKW and IWT\\
% ...
$ ^e$ Supported by the Deutsche Forschungsgemeinschaft \\
% ...
$ ^l$ Partially Supported by Russian Foundation
      for Basic Research, grant    no. 00-15-96584 \\
}

\end{flushleft}

\newpage

\section{Introduction}
\noindent
In QCD (Quantum Chromo Dynamics) the photoproduction of jets
with high transverse energy is 
described by the hard interaction of real photons 
with quarks and gluons inside the proton.
Interactions with two outgoing partons of large transverse momentum
are due to direct processes, such as
 $\gamma q \rightarrow gq$ (QCD-Compton effect)
and $\gamma g \rightarrow q\overline{q}$ (photon-gluon fusion) and
resolved processes where the photon first splits into a quark 
pair (or higher multiplicity fluctuation)
and one of the resulting partons subsequently scatters off a parton
in the proton.
The calculation of the latter processes can be approximated by
ascribing parton densities to the photon, which also include the inherently
non-perturbative aspects of the photon structure.

In analogy to the proton case parton densities of
the photon depend
on a factorization scale $\mu_\gamma$ and on $x_\gamma$, 
the longitudinal momentum
fraction of the photon taken by the interacting parton.
The limiting case
of direct interactions is given by $x_\gamma=1$. At HERA
these photoproduction reactions can be investigated in inelastic electron
(positron) proton reactions at very small squared four-momentum transfers
$Q^2$. Starting from the first investigation of this kind at HERA
\cite{Ahmed:1992xj}
the comparison of the predictions
of QCD with the results has been a central topic of
interest \cite{Adloff:1998da,Adloff:2000bs,Breitweg:1998rx}.
These investigations are particularly interesting, because
previous measurements of high transverse energy jet production in $ep$ and
$p\bar p$ scattering were not fully described by QCD calculations
\cite{Breitweg:1999wi,Affolder:2001ew,Abazov:2001hb}.

High transverse energy jets provide
a natural hard scale for perturbative QCD calculations.
Such calculations have been performed 
for direct and resolved processes in leading (LO) and next-to-leading
(NLO) order.
The measurement of jet cross sections at high transverse energy presented in this
paper can
therefore be used to test the current predictions of NLO
perturbative QCD and the parameterizations
of  photon and proton parton densities at
large scales with a precision of typically 10\%.
Photon quark densities have been determined
in experiments at  $e^+ e^-$-colliders\cite{Nisius:2000cv}
which investigate the photon structure function $F_2^\gamma$, where
$x_{\gamma}$ values up to $0.8$ and scales up  to $500 \gev^2$
have been reached.
In comparison the analysis presented here extends
the $x_{\gamma}$ range up to $1$  at
scales between $600$ and $6000 \gev^2$, where  the
quark density parameterizations of the photon
are presently not well constrained by measurements.
In contrast to the $F_2^\gamma$ measurements, the 
photoproduction of jets is
directly sensitive to the gluon density of the photon, which is 
poorly known to date.
Furthermore our data are sensitive to the parton densities of the proton at
fractional momentum values $x_p$ up to $0.6$.
In this kinematical regime, the quark densities
are well known from deeply inelastic scattering data, while the gluon
density has uncertainties of the order 10 to 50\%\cite{Huston:1998jj}.
Photoproduction data can thus be used to constrain
the parton density functions in regions where only few
measurements are presently available.
However, detailed parton densities can not be extracted 
from these data alone.

This paper
is based on an $e^+p$ data sample collected with the H1 detector
in the years 1995-1997 and corresponds to an 
integrated luminosity of 34.9 pb$^{-1}$.
It  presents dijet cross
sections as a function of jet observables with
mean jet transverse energies $20 < E_{T} < 80 \gev$,
observed $x_{\gamma}$ values $0.1 < x_{\gamma} < 1$
and values of $x_p$ ranging from
$0.05$ to $ 0.6$. 
The two jets considered in the investigated process $ep \rightarrow e\  {\rm jet}\,{\rm jet}\,X$ are
defined as the two jets with the highest transverse 
energy\footnote{A similar analysis has recently been made available\cite{ZEUSDiJet}.}.

\section{Jets in Photoproduction}
\label{sec:JetsinPhotoproduction}
\subsection{Cross sections and observables}
The cross section for the photoproduction of hard jets in electron-proton
collisions, $\sigma_{ep}$, can be calculated from the photon-proton scattering result, $\sigma_{\gamma p}$,
using the
factorization ansatz
\begin{equation}
  \sigma_{ep \rightarrow eX} =\int \dd y f_{ \gamma ,e}(y) \sigma_{\gamma p} (y)
  \enspace .
  \label{equ:sep}
\end{equation}
Here the usual variable $y$ of deeply inelastic scattering is interpreted
as the longitudinal momentum fraction of
the incoming electron taken by the photon and $f_{ \gamma,e}$ is
the  photon flux calculated in the Weizs\"acker-Williams
approximation\cite{wwa,Budnev:1974de,Frixione:1993yw}.
The hadronic photon-proton jet cross section is obtained
as the convolution
of the partonic cross sections
with the parton momentum distributions of the proton $f_{i/p}$ and the
photon $f_{j/ \gamma}$ .
As outlined in the introduction it is 
divided into a sum of two components, the
direct part $\sigma_{\gamma p}^{\mbox{\tiny{direct}}}$, where the 
photon
directly interacts with a parton of the proton and the resolved
part $\sigma_{\gamma p}^{\mbox{\tiny{resolved}}}$, where one
of the partons inside the photon interact
with a parton of the proton.
This distinction is  unambiguously defined in leading order only and depends
on the photon factorization scale $\mu_{\gamma}$. The two components
can be expressed as:

\begin{eqnarray}
 \sigma_{\gamma p}^{direct} & = &\sum_{i} \int \dd x_p f_{i/p}(x_p,\mu_{p}) \hat{\sigma}_{i\gamma}
 (\hat{s},\mu_{\gamma},\mu_p,\alpha_s(\mu_r),\mu_r) \\
  \sigma_{\gamma p}^{resolved} & = &\sum_{j,i} \int \dd x_{ \gamma} f_{j/ \gamma}
(x_{ \gamma},\mu_{\gamma}) \dd x_p f_{i/p}(x_p,\mu_{p}) \hat{\sigma}_{ij} 
(\hat{s},\mu_{\gamma},
\mu_p,\alpha_s(\mu_r),\mu_r)  
\enspace .\end{eqnarray}
The squared centre-of-mass energy of the hard subprocess is $\hat{s}=x_p x_{\gamma} y s$,
where
$\sqrt{s}$ is the total centre-of-mass energy in the $ep$-system, i.e. 300
GeV for this analysis.
The proton factorization scale is $\mu_p$ and the renormalization scale is
$\mu_r$.
The partonic cross sections $\hat{\sigma}$ can be expanded as a perturbative series
in powers of $\alpha_s$ and 
have been calculated up to the next-to-leading
order in QCD\cite{Frixione:1997ks,Klasen:1997it,Harris:1997hz,Aurenche:2000nc}.

The total cross sections
on the left hand side of equation~\ref{equ:sep} are
obtained by integrating over $y$, $x_p$ and $x_\gamma$. The partonic
cross sections $\hat{\sigma}_{i\gamma}$ and $\hat{\sigma}_{ij}$
contain a further integration over an internal
degree of freedom, e.g. $\cos \theta^*$, the scattering angle
in the centre-of-mass system of the partonic two body reaction, or
the transverse energy.
More detailed information on the reaction dynamics is 
obtained by measuring differential cross sections in these kinematical
observables.
In order to avoid singularities in the partonic cross sections
a minimum cut in $\theta^*$ or in the transverse energy of the
outgoing partons has to be applied.

The two scaled longitudinal parton momenta $x_{\gamma}$ and $x_{p}$
are calculated
from the jets produced in the hard subprocess, using the
definition
\begin{eqnarray}
 x_{\gamma} & = & \frac{1}{2 E_e y} (E_{T,1}
e^{- \eta_1} + E_{T,2} e^{- \eta_2})\label{xgam} \label{equ:xg}\\
 x_{p} & = & \frac{1}{2 E_p} (E_{T,1} e^{ \eta_1} + E_{T,2} e^{ \eta_2})
\label{equ:xp}
\end{eqnarray}
Here $E_{T,1}$ and $E_{T,2}$ are the transverse energies of the two
jets of the hard subprocess, $\eta_1$ and $\eta_2$ are their pseudorapidities 
in the laboratory frame ($\eta=-\ln (\tan \theta/2$)) and
$E_e$ and $E_p$ are the energies of the electron and proton beams\footnote{The
coordinate system is centered at the nominal
interaction point with
the positive $z$ direction along the incident proton beam.
The polar angle $\theta$ is defined with respect to the positive
$z$ axis.}.
These relations are used as definitions of observables in all orders
and are easily derived for $2\rightarrow 2$ processes,
where the transverse energies of the jets are equal.
The pseudorapidities of the jets are related to $\theta^*$
via
\begin{equation}
\cos \theta^* = | \tanh ((\eta_1-\eta_2)/2) | \enspace .
\end{equation}

In principle one could measure the dependence of the
fourfold differential cross section
$d\sigma_{\gamma p}/dydx_\gamma dx_p d\cos\theta^*$ on all
four variables.
This, however, would require a much larger data set than
presently available.
Therefore in this paper
more inclusive quantities are presented.
The distribution of the invariant mass of the two jets
with the highest transverse energies, $M_{JJ}$,
the mean transverse energy of the two leading jets $E_{T,mean}$
and the transverse energy distribution
of the highest transverse energy jet, $E_{T,max}$, are studied.
The cross section differential in
the average value of the pseudorapidities
$\overline{\eta}=(\eta_1+\eta_2)/2$ is particularly sensitive to
parton density functions. It is thus presented  for different
photon-proton centre-of-mass energies ($y$ regions) and different scales
($E_{T,max}$ regions), cf. equations~\ref{equ:xg} and \ref{equ:xp}.

Differential cross sections in  $x_{\gamma}$ and $x_{p}$
are measured
in different scale regions ($E_{T,max}$ regions)
and for different $x_{\gamma}$ or $x_p$
cut-off values.
The angle $\theta^*$ is sensitive to the dynamics of jet production and
the corresponding differential cross section  is therefore evaluated
for different
$x_{\gamma}$ regions for all $M_{JJ}$ and in addition with a cut 
in $M_{JJ}$.
The $\cos\theta^*$ distribution could be influenced by the production
of $W$ or $Z^0$ bosons, whose hadronic decays  
have a different angular distribution from that expected for
QCD dijet production.
Using the EPVEC Monte Carlo generator \cite{Baur:1992pp}, 
the contribution of $W$ bosons
is estimated to be 5-6 events. The $Z^0$ contribution  
is expected to be negligible. The background from these processes is
therefore not considered in the following.

In the present analysis
jets are defined using the inclusive $k_{\bot}$ algorithm
as proposed in \cite{Ellis:1993tq,Catani:1993hr}. The application of this
algorithm has become standard in jet analyses at HERA \cite{Adloff:2001tq}.
It utilizes a definition of jets in which not
all particles are assigned to hard jets.
Here it is applied in the laboratory frame with the separation parameter
set to 1 and using
an $E_T$ weighted recombination scheme in which the jets are considered massless.

\subsection{QCD Predictions and Models}
To simulate the direct and resolved  photoproduction of jets, the
PYTHIA 5.7~\cite{Sjostrand:1994yb} and HERWIG 5.9~\cite{Marchesini:1992ch} event generators
were used followed by
a full detector
simulation~\cite{Brun:1987ma} of all Monte Carlo events.
Both programs contain the Born level QCD hard scattering
matrix elements, regulated by a minimum cut-off in transverse momentum. 
Leading
logarithmic parton showers are used to represent higher order QCD radiation.
GRV-LO~\cite{Gluck:1994uf,Gluck:1992jc} 
parton density functions (pdfs) for
the proton and photon were chosen. The Lund String model is applied in
PYTHIA to hadronize the outgoing partons, while in HERWIG the
cluster hadronization approach is used. Multiple interactions between the
proton and the resolved photon are dealt with in PYTHIA by adding
additional interactions
between spectator partons within the same event.
These processes are calculated by extending the
perturbative parton-parton scattering to a low $E_T$ cut-off.

In HERWIG  multiple interactions are included 
by producing in a fraction $P'$ of the resolved events
so called {\it soft underlying events}. These interactions 
are parameterized using experimental results of soft hadron-hadron
scattering.  
The effect of multiple interactions is tested by comparing,
in the data and in the HERWIG calculations, the 
energy flow distributions around the jet axis 
with and without a fraction $P'$ of 
events containing the soft underlying
event.
For $P'$ $\sim$30-35\% these distributions are found to be
well described for all regions of $x_{\gamma}$.
The difference of the calculated HERWIG
cross sections with and without $35$\% of soft underlying events
is below 10\% for $x_{\gamma}$ between 0.3-0.8 and
10-20\% for $x_{\gamma} < 0.3$.
For $x_{\gamma}>0.8$ the difference is negligible.
PYTHIA is also able to describe these distributions.

The goal of this analysis is the comparison of the
measured cross sections to perturbative QCD
calculations at the parton level.
The LO and NLO dijet cross sections were computed using a program based
on the subtraction method \cite{Frixione:1997ks,Frixione:1997np}
for the analytic cancellation of infrared singularities. In calculating
LO and NLO cross sections a 2-loop $\alpha_s$ was taken with 5 active
flavours.
$\Lambda_{QCD}$ was set to $0.226$ GeV ($\alpha_s(M_Z)=0.118$), 
which is the value used
in the proton parton density functions.
CTEQ5M~\cite{Lai:1997mg} parton density functions were chosen for the proton
whereas
MRST99~\cite{Martin:2000ww} parton density functions were selected to test the
dependence of the NLO cross sections on the proton pdfs.
For the photon we  choose GRV-HO~\cite{Gluck:1992ee} as a main setting and the
parameterization of AFG-HO~\cite{Aurenche:1994in} to study
the dependence of the results on the choice of the photon pdfs.
The  renormalization scale $\mu_r$ and the factorization scales
$\mu_p$ and $\mu_\gamma$ were, event by event, 
set to the sum of the transverse
energies of the outgoing partons divided by two.
The QCD program allows the variation of this common scale.
It was varied
from 0.5 to 2 times the default scale to estimate
the scale uncertainty in the NLO calculation.
This uncertainty turned out to vary between $\pm 10$ and $\pm 20\%$
in the measured kinematic range.

In addition the data are compared to the predictions of NLO QCD
corrected for hadronization effects, which
are defined as the ratio
of the cross sections with jets reconstructed from  hadrons
and from partons before hadronization.
The hadronization effects are calculated with 
PYTHIA and HERWIG and the mean value of the
two predictions is used for corrections.
Here the difference between the two Monte Carlo models
is in general very small and
at maximum 10\%. The jets built out of partons are found to
be very well
correlated with the jets built out of hadrons.

\section{Experimental Technique}
\subsection{H1 Detector}
The H1 detector is described in detail in~\cite{Abt:hi,Abt:1996xv}. Only those
components relevant to the present analysis are briefly described here.
The Liquid
Argon (LAr )~\cite{Andrieu:1993kh} and SpaCal~\cite{Appuhn:1996na} 
calorimeters were used
to trigger events, to reconstruct the hadronic energy of the final state
and to select photoproduction events by eliminating
events with an identified scattered positron.
The LAr calorimeter covers the polar angle range $4^\circ < \theta <
154^\circ$ with full azimuthal acceptance.
The jet energy calibration agrees at the 2\% level
with the Monte Carlo simulation 
as
determined by the transverse energy balance between jet 
and electron for deeply inelastic scattering
events and by the transverse energy balance between the two jets 
for the photoproduction
sample in different kinematic regions.
The angular region
\mbox{$153^\circ < \theta < 177.8^\circ$} is covered by the SpaCal, a
lead/scintillating-fibre calorimeter.
It has a hadronic energy scale
uncertainty of 8\%.
The central tracking detector (CJC) was used to
reconstruct the interaction vertex and to supplement the measurement
of hadronic energy flow.
The CJC consists of two concentric cylindrical drift chambers, coaxial
with the beam-line, with a polar angle coverage of $15^\circ <
\theta < 165^\circ$. The entire CJC is immersed in a 1.15 T magnetic
field. The luminosity determination is based on the measurement of the $ep
\rightarrow ep \gamma$ Bethe-Heitler process, where the positron and
photon are detected in calorimeters located
downstream of the interaction point in $e$-beam direction.

\subsection{Event Selection}
The data sample was collected at HERA with the H1 detector in the 
years 1995-97, when
protons of 820 GeV energy collided with positrons of 27.6~GeV energy 
resulting in a
centre-of-mass energy of 300~GeV.
The events were triggered on the basis of high transverse energy
deposits in the LAr calorimeter.
The trigger efficiencies were above 94\% for the
event sample described in this analysis.
Energy deposits in the calorimeters and tracks
in the CJC were combined in a manner that avoids double
counting to reconstruct the hadronic energy of
events~\cite{Adloff:1997mi}. 

It was required that an event vertex
was reconstructed within
$35$ cm of the nominal $z$ position of the vertex.
The most significant background in the data sample arises from
neutral current deeply inelastic scattering events, and
was suppressed by removing events with an electron
identified in the LAr calorimeter or SpaCal and by
requiring $y<0.9$, with $y$ reconstructed using hadronic
variables~\cite{yJB}.
This reduces the 
background to less than 1\% for the total sample. 
In the region with the highest $y$ at low $\overline{\eta}$ the
remaining background was calculated
to be about 5\% based on deeply inelastic scattering dijet data
and the ARIADNE~\cite{Lonnblad:1992tz}
Monte Carlo interfaced with DJANGO~\cite{Charchula:1994kf}. 
It was subtracted statistically.
After applying a cut on the missing transverse energy
$E_{T,miss}<$ 20 GeV the remaining  charged current 
($ep \rightarrow \nu X$) and non-$ep$
scattering background was found to be negligible.  
Events induced by cosmic rays were removed.

Asymmetric cuts on
the $E_T$ of the two jets with the highest transverse energies
are applied to avoid regions of phase space affected by 
uncertainties in the NLO calculation\cite{Frixione:1997ks}.
On the other hand a highly asymmetric cut 
causes large NLO corrections and a pronounced dependence on the
choice of scale.
The jet selection criteria therefore required an $E_T$ of the highest
transverse energy jet $E_{T,max}>25$~GeV,
and the transverse energy of the second highest
transverse energy jet $E_{T,second}>15$~GeV.
When the cut on $E_{T,second}$ is varied between $\pm 5$~GeV,
the ratio of the measured cross sections to the theoretical prediction
varies by up to 10\% for $x_{\gamma}<0.8$ and by up to 3\%
for $x_{\gamma}>0.8$.

\begin{table}[ht]
\begin{center}
\begin{tabular}{|c|}
\hline
$Q^2<1 \mbox{ GeV}^2$\\
$0.1<y<0.9$ \\
$E_{T,max}>25$~GeV\\
$E_{T,second}>15$~GeV\\
$-0.5 < \eta_{i} < 2.5$ \\
\hline
\end{tabular}
\caption{The definition of the phase space of the measured dijet cross sections.\label{tab:phasespace}} 
\end{center}
\end{table}

The
pseudorapidity of each jet $\eta_i$ was restricted to $-0.5 < \eta_{i} <
2.5$. All jets are thus well contained in the LAr calorimeter.
The measured kinematic
region was restricted to $0.1< y < 0.9$ and $Q^2 < 1 \gev^2$, as
given by the acceptance for electrons in the LAr and Spacal.
The kinematic range of the measured dijet cross sections
is summarized in Table~\ref{tab:phasespace}.
Applying these cuts the total number of events measured was  5265.

\subsection{Correction of the Data for Detector Effects}
The data were corrected for detector effects such as limited resolution
and inefficiencies.
To determine these effects the HERWIG and PYTHIA Monte Carlo
samples were used.
Both programs do not describe the absolute normalization of 
the dijet cross sections. 
After scaling the HERWIG cross sections by $2$ and the PYTHIA 
cross sections by $1.2$ the two 
programs gave a good description of the 
measured jet distributions in 
shape and normalization.

The bin sizes of all distributions are 
matched to the resolution and result in 
good bin
efficiency and purity.
The correction was done by using the so called bin-to-bin method.
The correction functions were calculated from the ratio
of the cross sections with jets 
reconstructed from hadrons (hadron level) and
from detector objects (detector level)
in each bin, where each sample was subject to the selection
criteria defined above.
The correction functions of the two models are in good agreement
and differ on average by 5\% and at most by 20\%.
The mean values of the two Monte Carlo generators were thus taken
for the correction.
The resulting correction factors typically have values between
$0.8$ and $1.2$.

\subsection{Systematic Uncertainties}
For the jet cross sections the following sources of systematic error were considered:
\begin{itemize}
\item A 2\% uncertainty in the LAr energy scale
  results in an uncertainty of typically
  10\%.
\item An 8\% uncertainty in the hadronic Spacal energy scale
  results in an uncertainty of 1\%.
\item
  In addition to the variations of the calorimeter energy
  scales a shift
  of 1\% on $y$ is considered.
  This variation results in an uncertainty of
  3\%.
\item
  Half of the difference between the correction factors
  calculated with HERWIG and with PYTHIA
  is taken as the uncertainty in the detector correction.
  The resulting uncertainty is less than 10\%.
\item The uncertainty in the trigger efficiency results in an error of
  $\sim 3$\%.
\item The uncertainty in the background subtraction results in
  an error
  of $\sim 2$\%.
\item The uncertainty in the integrated luminosity results in an
  overall normalization error of 1.5\%.
\end{itemize}

The statistical 
and all systematic errors are added in quadrature.
The resulting total uncertainty ranges from 10 to 30\%, where
the systematic contribution is dominated by uncertainties in the
calorimeter energy scales and in the correction to the hadron level.

\section{Results}

The measured cross sections for inclusive dijet production
in the reaction $ep \rightarrow e\  {\rm jet}\,{\rm jet}\,X$ are given
as single differential cross sections in all cases.
The data are corrected for detector effects and are presented 
at the level of stable hadrons for
the phase space region defined in Table~\ref{tab:phasespace}.
The inner
error bars of the data points in the figures denote the statistical,
the outer error bars the total uncertainty.
The data are also presented in Tables~\ref{tab:MJJ}-\ref{tab:costh}.
All results are compared to next-to-leading order (NLO) QCD
predictions obtained with the standard setting described in section~\ref{sec:JetsinPhotoproduction}
if not otherwise quoted.
The predictions of NLO QCD
corrected for hadronization effects NLO$(1+\delta_{hadr})$ are also shown.

\begin{figure}[p]
\center
\epsfig{file=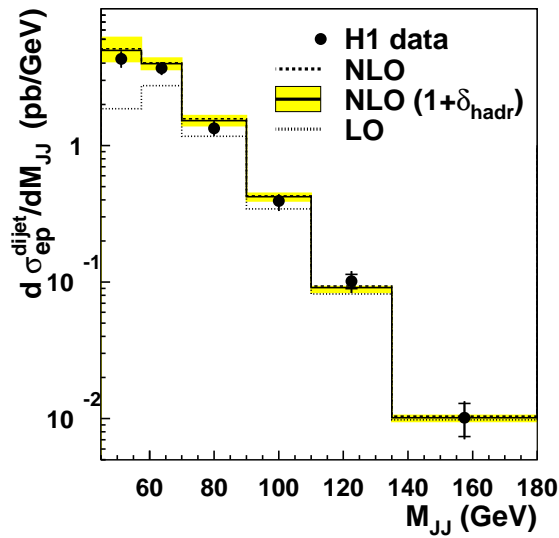,width=0.5\textwidth}
\caption{Differential $ep$ cross sections for dijet production 
  ($Q^2< 1\mbox{GeV}^2$) as a function 
  of
  the invariant dijet mass $M_{JJ}$ of the two highest $E_T$ jets.
  Here, as well as in the following figures unless explicitly stated otherwise,
  the inner
  error bars denote the statistical error, the outer error bars the
  total uncertainties of the data.
  The LO predictions
  using CTEQ5M pdfs for the proton and
  GRV-HO pdfs for the photon are
  shown as a dotted line.
  NLO predictions with the same pdfs are shown
  as a dashed line.
  The full line shows the NLO predictions, including hadronization corrections  and the grey band
  indicates the renormalization and factorization
  scale uncertainties of the NLO prediction.\label{fig:fig1}}

\end{figure}

In Figure~\ref{fig:fig1} the dijet cross section is shown as a function of
the invariant mass $M_{JJ}$ of the dijet system.
The data are presented for $M_{JJ}$ values between $45$ and
$180$ $\gev$. The measured cross section falls by about
3 orders of magnitude over this range.
NLO QCD describes the measured cross sections for the
whole mass range.
Hadronization corrections are less than 5\% for all bins.
The calculation using LO matrix
elements fails to describe the low
$M_{JJ}$ region. This is partly due to the fact that the
low $M_{JJ}$ region is populated by events which are influenced
by the asymmetric cuts on the jet transverse energies.
Events in which the second jet has a transverse energy below $25$ $\gev$
contribute mainly in this region.
In dijet calculations they only appear beyond leading order.
The scale uncertainties
in the QCD predictions are largest at low $M_{JJ}$ values.

\begin{figure}[p]
\center
\epsfig{file=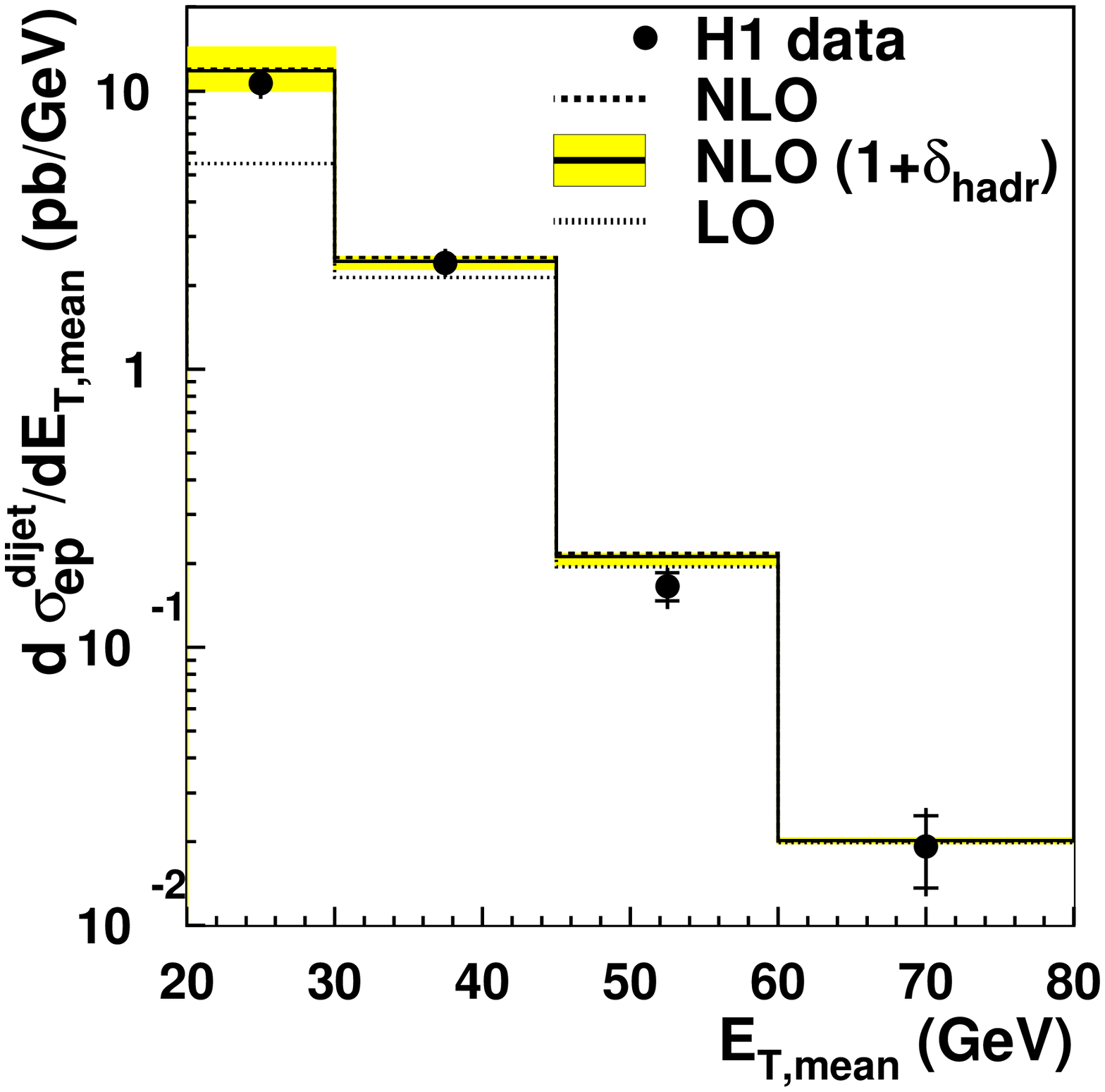,width=0.5\textwidth}
\put(-60,20){\Large{a)}}
\epsfig{file=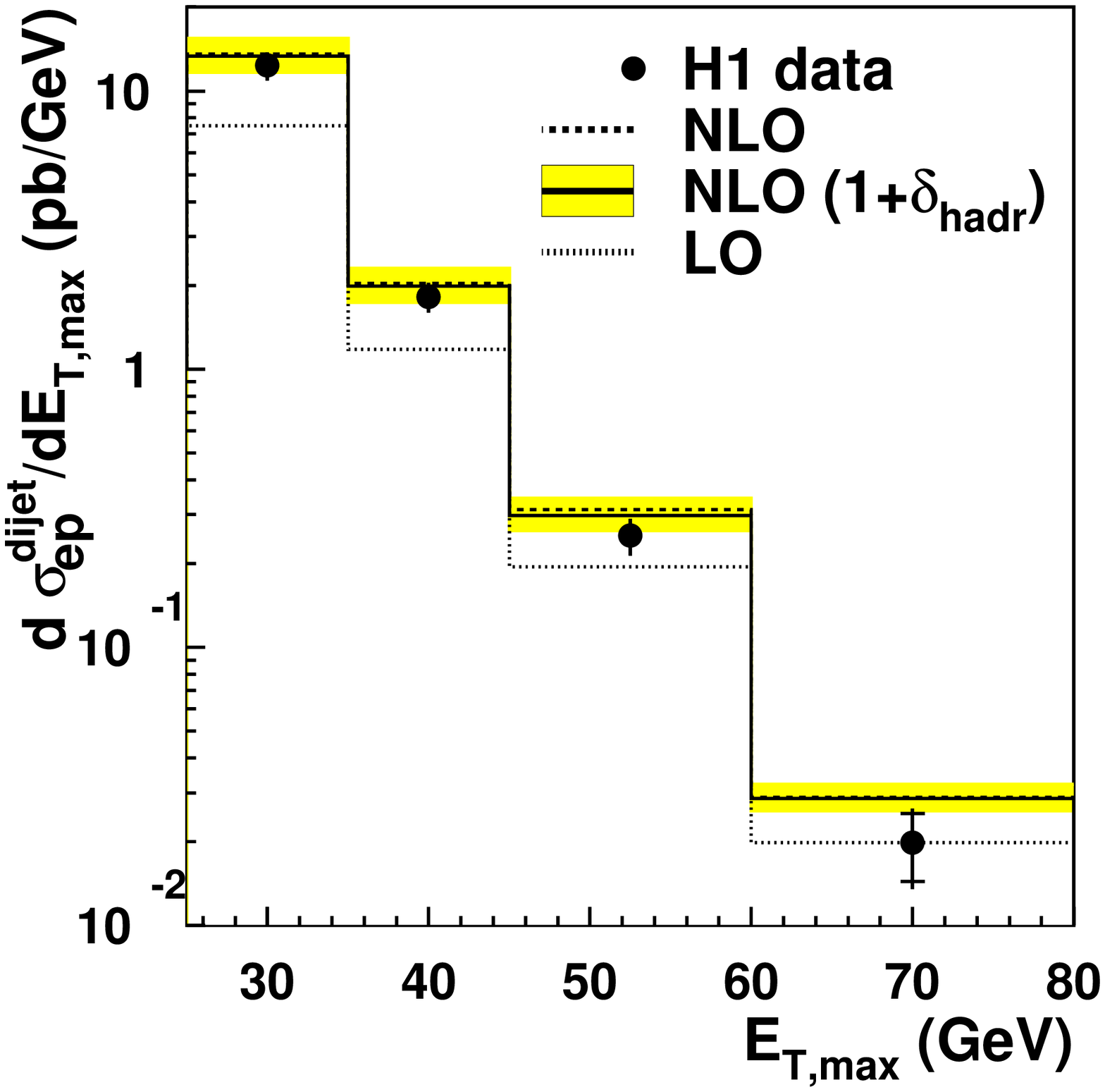,width=0.5\textwidth}
\put(-60,20){\Large{b)}}

\caption{Differential $ep$ cross sections for dijet production 
  ($Q^2< 1\mbox{GeV}^2$) as a function 
  of a) $E_{T,mean}$, the mean
  and b) $E_{T,max}$, the maximum
  $E_T$ of the two highest $E_T$ jets.
  The LO predictions
  using CTEQ5M pdfs for the proton and
  GRV-HO pdfs for the photon are
  shown as a dotted line.
  NLO predictions with the same pdfs are shown
  as a dashed line.
  The full line shows the NLO predictions, including hadronization corrections  and the grey band
  indicates the renormalization and factorization
  scale uncertainties of the NLO prediction.\label{fig:fig2}}

\end{figure}

A similar statement on the large scale uncertainties 
and the difference between data and the LO calculation holds for small transverse momenta.
In Figure~\ref{fig:fig2}a) the 
dijet cross section
$d\sigma / d E_{T,mean}$ is shown.
Here the scale uncertainties decrease from $\pm 20$\% 
for the first bins to less than $\pm 5$\% for $E_{T,mean}>30$ GeV.
The data are well described by the NLO calculation.
The dijet cross section as a function of the transverse energy
of the highest transverse energy jet $E_{T,max}$
is shown in Figure~\ref{fig:fig2}b).
The distribution again demonstrates that the data are
described by
NLO QCD up to the highest $E_{T,max}$ values within errors.
The NLO scale uncertainty is not reduced significantly
with increasing $E_{T,max}$.
The cross sections differential in transverse energy 
are hardly altered by
hadronization corrections which are around 5\% for all bins.
The NLO QCD calculation with hadronization corrections predicts 
the measured cross sections up to the highest masses and
transverse energies, although the photon and proton pdfs
have been 
extracted from quite different processes and mostly at lower 
scales.

\begin{figure}[p]
\center
\epsfig{file=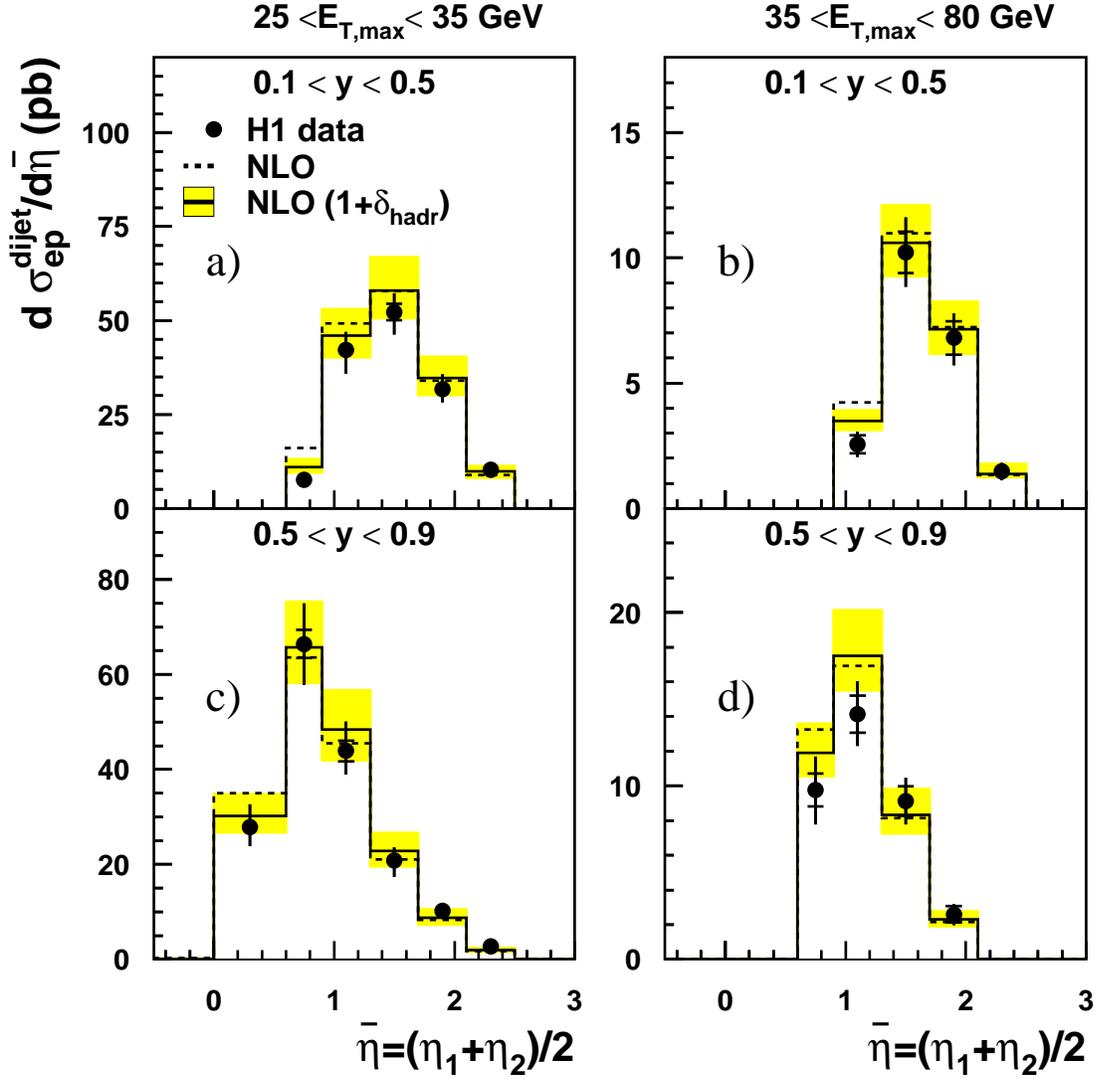,width=\textwidth}
\put(-133,123){\Large{a)}}
\put(-65,123){\Large{b)}}
\put(-133,65){\Large{c)}}
\put(-65,65){\Large{d)}}
\caption{Differential $ep$ cross sections for dijet production 
  ($Q^2< 1\mbox{GeV}^2$) as a function 
  of
  the $\overline{\eta}$ of the two highest $E_T$ jets.
  The regions of low $y$, a) and b) and high $y$, c) and d) are shown for two
  ranges of $E_{T,max}$.
  NLO predictions
  using CTEQ5M pdfs for the proton and
  GRV-HO pdfs for the photon are
  shown as a dashed line.
  The full line shows the NLO predictions, including hadronization corrections  and the grey band
  indicates the renormalization and factorization
  scale uncertainties of the NLO prediction.\label{fig:fig3}}

\end{figure}

To further explore the photon and proton structure the
differential
cross section $d\sigma / d\overline{\eta}$
is displayed in Figure~\ref{fig:fig3} 
for two ranges of $E_{T,max}$ subdivided into two $y$
regions. While the former implies a variation of the scale 
the latter corresponds to different center of mass
energies in the photon-proton-system. 
Again, good agreement between data and NLO QCD is observed 
taking into account the uncertainties in
the calculations and in the data points. 
The predictions tend to lie above
the data at low $\overline{\eta}$, where direct interactions
dominate and hadronization corrections are largest. 
At high $\overline{\eta}$, where in contrast resolved interactions 
dominate and hadronization corrections are small,
the NLO QCD predictions agree well with the
measured data.

\begin{figure}[p]
\center
\epsfig{file=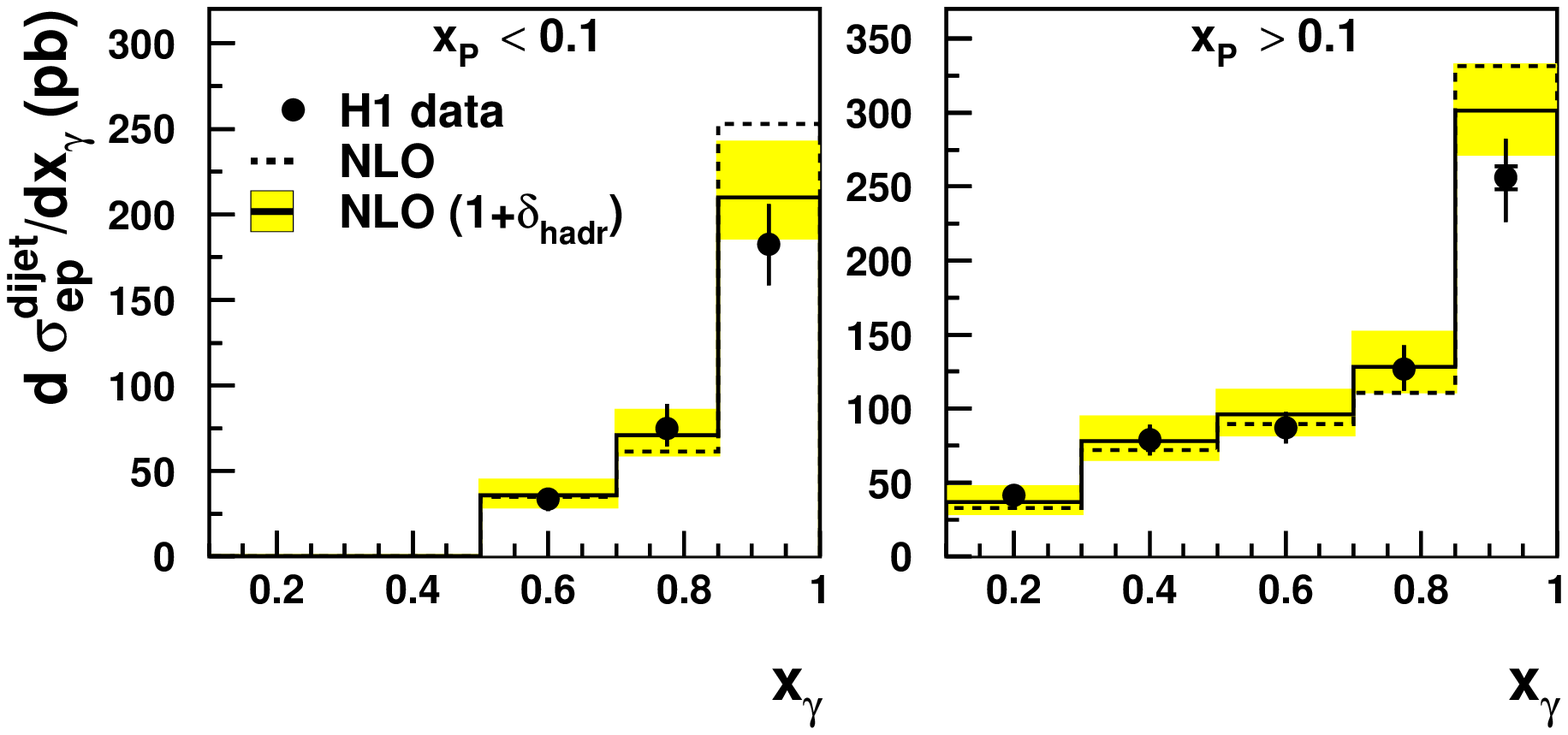,width=\textwidth}
\put(-135,45){\Large{a)}}
\put(-65,45){\Large{b)}}

\epsfig{file=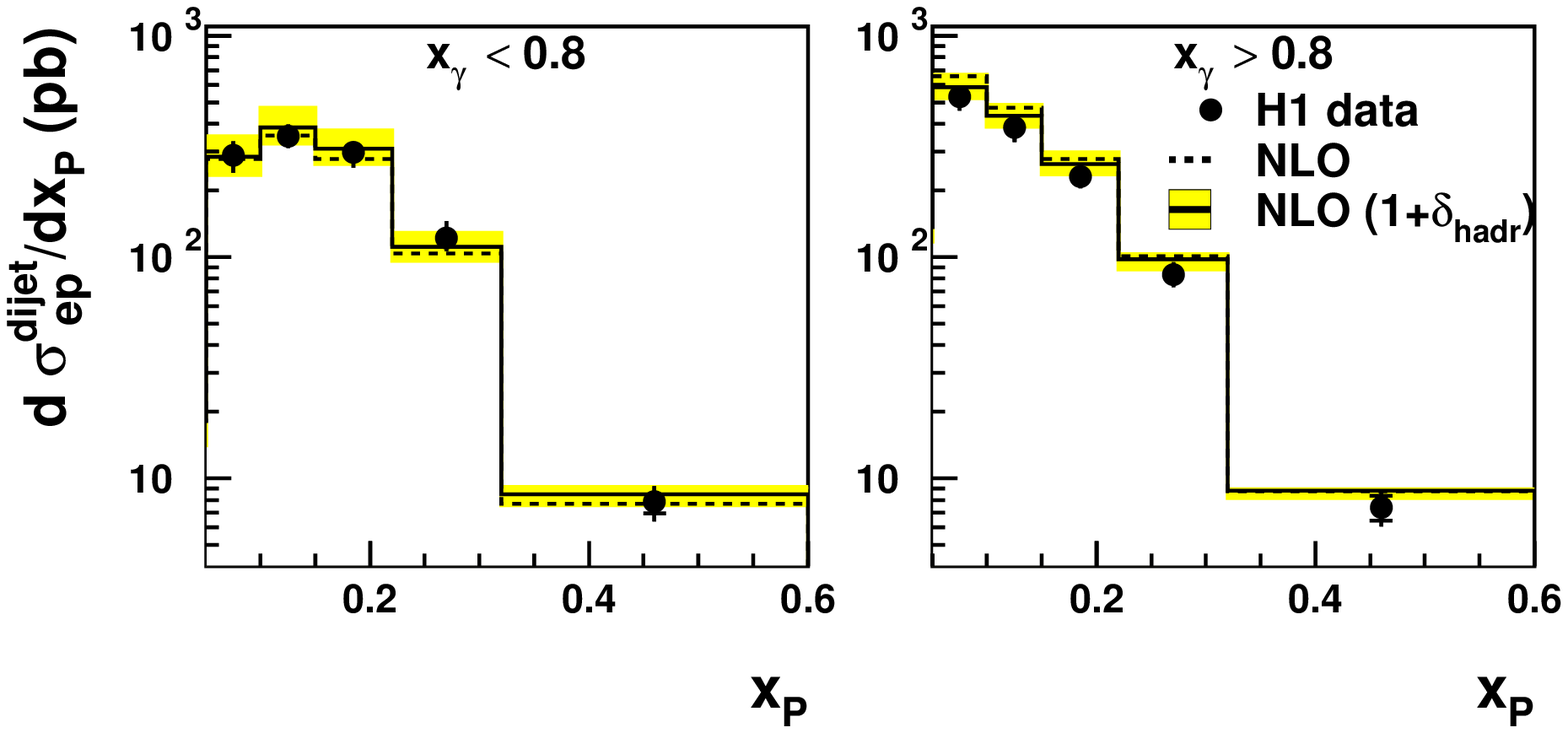,width=\textwidth}
\put(-135,45){\Large{c)}}
\put(-65,45){\Large{d)}}

\caption{Differential $ep$ cross sections for dijet production 
  ($Q^2< 1\mbox{GeV}^2$) as a function of
  $x_{\gamma}$ a) and b) and $x_p$ c) and d).
  Figures a) and b) distinguish regions of small and large $x_p$
  and figures c) and d) corresponding regions in $x_{\gamma}$.
  NLO predictions
  using CTEQ5M pdfs for the proton and
  GRV-HO pdfs for the photon are
  shown as a dashed line.
  The full line shows the NLO predictions, including hadronization corrections  and the grey band
  indicates the renormalization and factorization
  scale uncertainties of the NLO prediction.\label{fig:fig4}}

\end{figure}

Figures~\ref{fig:fig4}a) and b) show the dijet cross section
$d \sigma / d x_{\gamma}$
as a function of
$x_{\gamma}$
for two different $x_{p}$ regions.
The calculations exceed the data,
while remaining within the given uncertainties, only for
$x_{\gamma}>0.85$, where the largest hadronization
correction occur.
Using the MRST99 1-3 proton pdfs (with a large variation of the high $x_p$ gluon density) instead of CTEQ5M results in differences
of less than 5\% for the predicted cross section
for $x_p < 0.1$ and up to 15\% for $x_p > 0.1$.
This is smaller
than the scale uncertainties for $x_p < 0.1$ and
of the same order for $x_p > 0.1$.
These findings are corroborated in Figures~\ref{fig:fig4}c) and d) where
the cross section $d\sigma / d x_p$ is shown 
as a function of $x_p$
for two different $x_{\gamma}$ regions.
Even at the highest $x_p$ the measured cross sections
are seen to agree well with the QCD predictions, which
in this part of the phase space attribute about 40\% of the 
cross section to processes induced by gluons in the proton.
The constraints on the pdfs used in the QCD calculations here come dominantly from deeply inelastic scattering at lower scales where the gluon fraction is smaller.
The concept of universal pdfs in hard processes in QCD is thus observed to describe measurements with rather different experimental conditions.

\begin{figure}[p]
\center
\epsfig{file=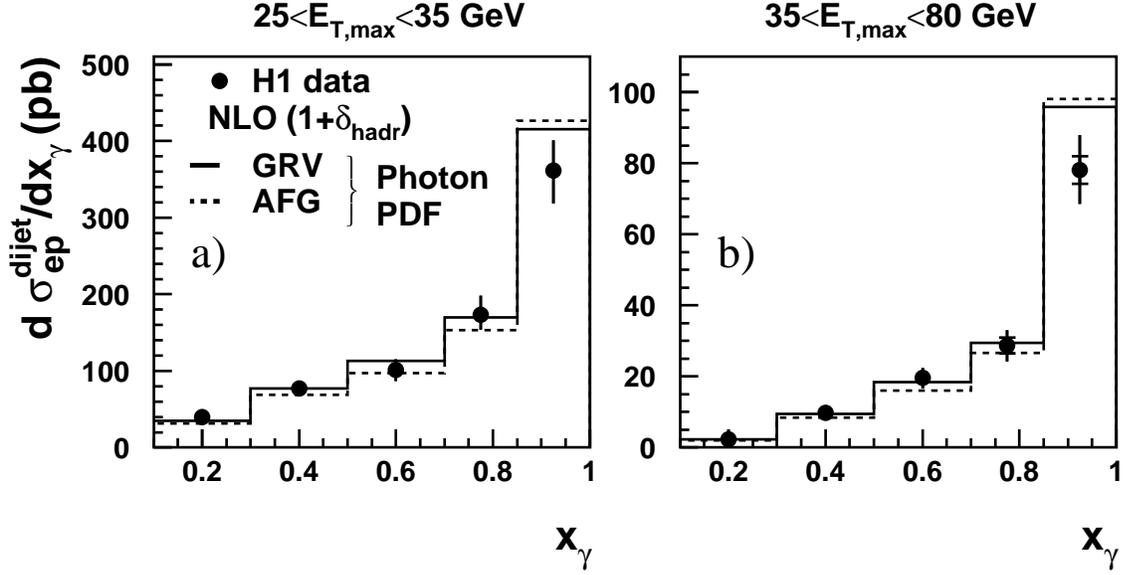,width=\textwidth}
\put(-135,45){\Large{a)}}
\put(-65,45){\Large{b)}}
\caption{Differential $ep$ cross sections for dijet production 
  ($Q^2< 1\mbox{GeV}^2$) as a function of
  $x_{\gamma}$ for a) low $E_{T,max}$
  and b) high $E_{T,max}$.
  The NLO predictions
  using CTEQ5M pdfs for the proton and
  GRV-HO pdfs for the photon and
  including hadronization corrections
  are shown
  as a full line. NLO predictions using AFG-HO parametrizations
  of the photon pdfs and including hadronization
  corrections are shown as the dashed line.\label{fig:fig5}}

\end{figure}

\begin{figure}[p]
\center
\epsfig{file=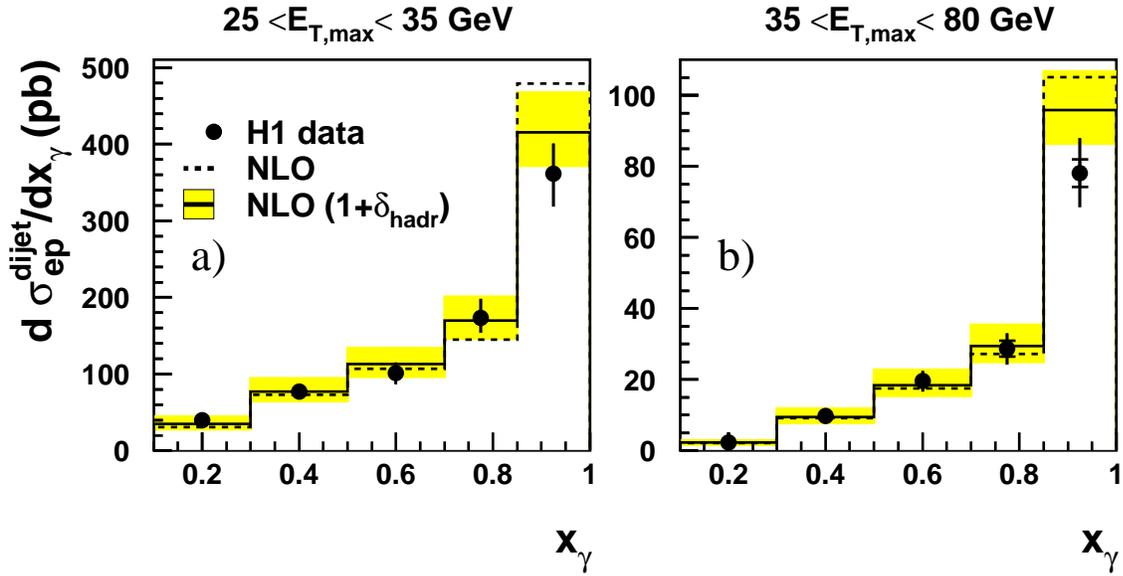,width=\textwidth}
\put(-135,45){\Large{a)}}
\put(-65,45){\Large{b)}}
\caption{Differential $ep$ cross sections for dijet production 
  ($Q^2< 1\mbox{GeV}^2$) as a function of
  $x_{\gamma}$ for a) low $E_{T,max}$
  and b) high $E_{T,max}$.
  NLO predictions
  using CTEQ5M pdfs for the proton and
  GRV-HO pdfs for the photon are
  shown as a dashed line.
  The full line shows the NLO predictions, including hadronization corrections  and the grey band
  indicates the renormalization and factorization
  scale uncertainties of the NLO prediction.\label{fig:fig6}}

\end{figure}

\begin{figure}[p]
\center
\epsfig{file=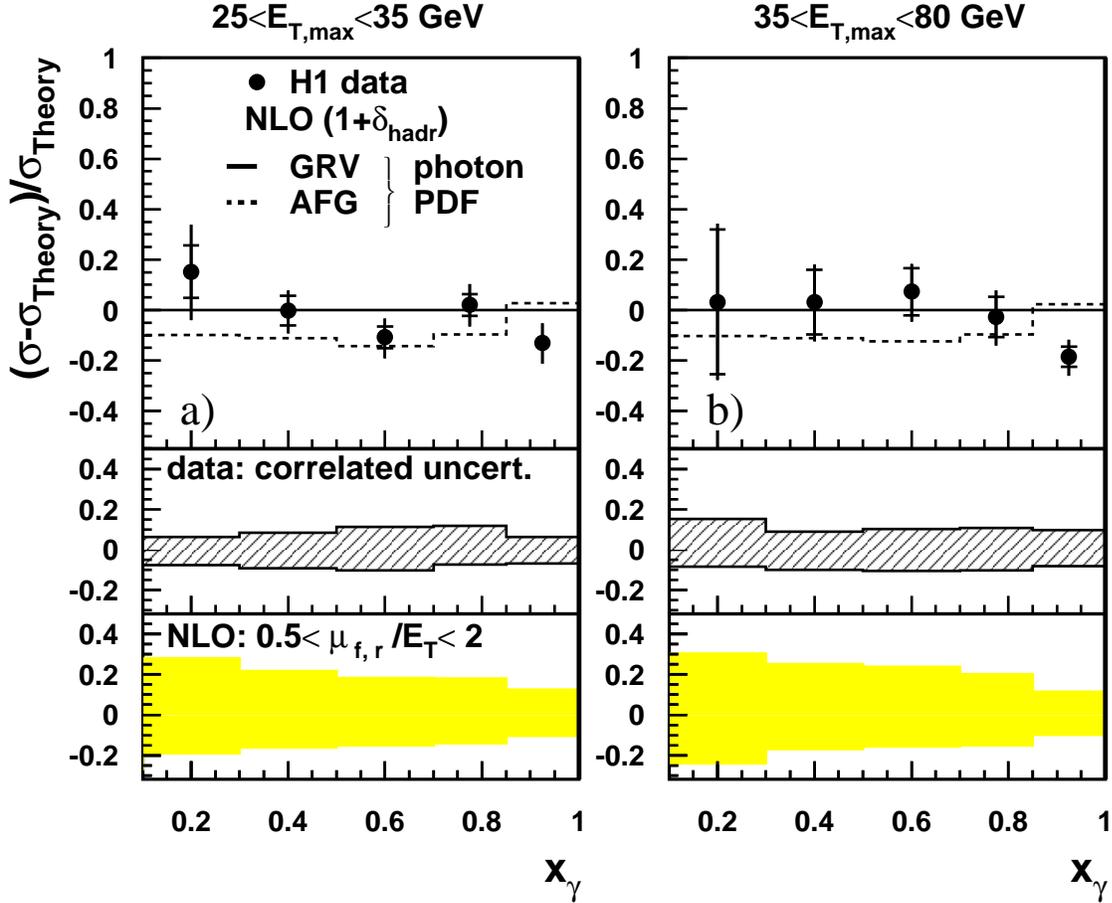,width=\textwidth}
\put(-135,75){\Large{a)}}
\put(-65,75){\Large{b)}}
\caption{The $x_{\gamma}$ dependence of the relative 
difference of the measured dijet cross sections ($Q^2< 1\mbox{GeV}^2$)
 from 
 the NLO prediction, with hadronization corrections applied
 using CTEQ5M pdfs for the proton and
  GRV-HO pdfs for the photon (here $\sigma_{Theory}$).
  The symbol $\sigma$ stands for $d \sigma / d x_{\gamma}$.
  Shown is the relative difference of 
  the data (points) and the NLO predictions
  using the AFG-HO pdf (dashed line)  
  with  hadronization corrections applied.
  Figures a) and b) show the relative difference
  for the lower $E_{T,max}$
  and higher $E_{T,max}$ regions respectively.
  The inner
  error bars denote the statistical error, the outer error bars
  denote  
  all statistical and uncorrelated
  systematic errors of the data added in quadrature.
  The correlated systematic errors are shown in the middle plots
  as a shaded band.
  The grey band (lower plots)
  shows the renormalization and factorization
  scale uncertainties of this NLO prediction.\label{fig:fig7}}
\end{figure}

Figure~\ref{fig:fig5} displays the dijet cross sections $d \sigma / d x_{\gamma}$
as a function of $x_{\gamma}$ for
two regions of $E_{T,max}$, representing
different factorization scales for the photon and proton pdfs.
The data are compared to NLO calculations corrected for
hadronization effects
with two different parameterizations of the photon structure.
The predictions describe the data well and 
vary only slightly with the photon pdfs used.
In contrast the NLO scale uncertainties produce a significant
effect as can be inferred from 
Figure~\ref{fig:fig6}, which repeats the data of Figure~\ref{fig:fig5} with
a comparison of the GRV-HO pdfs of the photon. 
For high values of $x_{\gamma}$ the hadronization 
corrections are sizeable and improve the agreement with the data.
A more detailed comparison between data and theory
is obtained by plotting their relative
difference as shown in Figure~\ref{fig:fig7}.
NLO predictions including hadronization
corrections are shown for both sets of photon pdfs.
At variance to the previous plots the error bars of the data
contain only the uncorrelated systematic errors, while the correlated
errors due to the uncertainty in the calorimeter energy scales are
shown as a hatched band.
Figure~\ref{fig:fig7} shows 
that the assumed NLO scale uncertainties are the dominant source 
of uncertainties in the comparison of data and theory.
The expectation exceeds 
the data only for the high $x_{\gamma}$ and 
high $E_{T,max}$ regions.
Within these uncertainties the picture of an universal photon structure is thus corroborated.

\begin{figure}[p]
\center
\epsfig{file=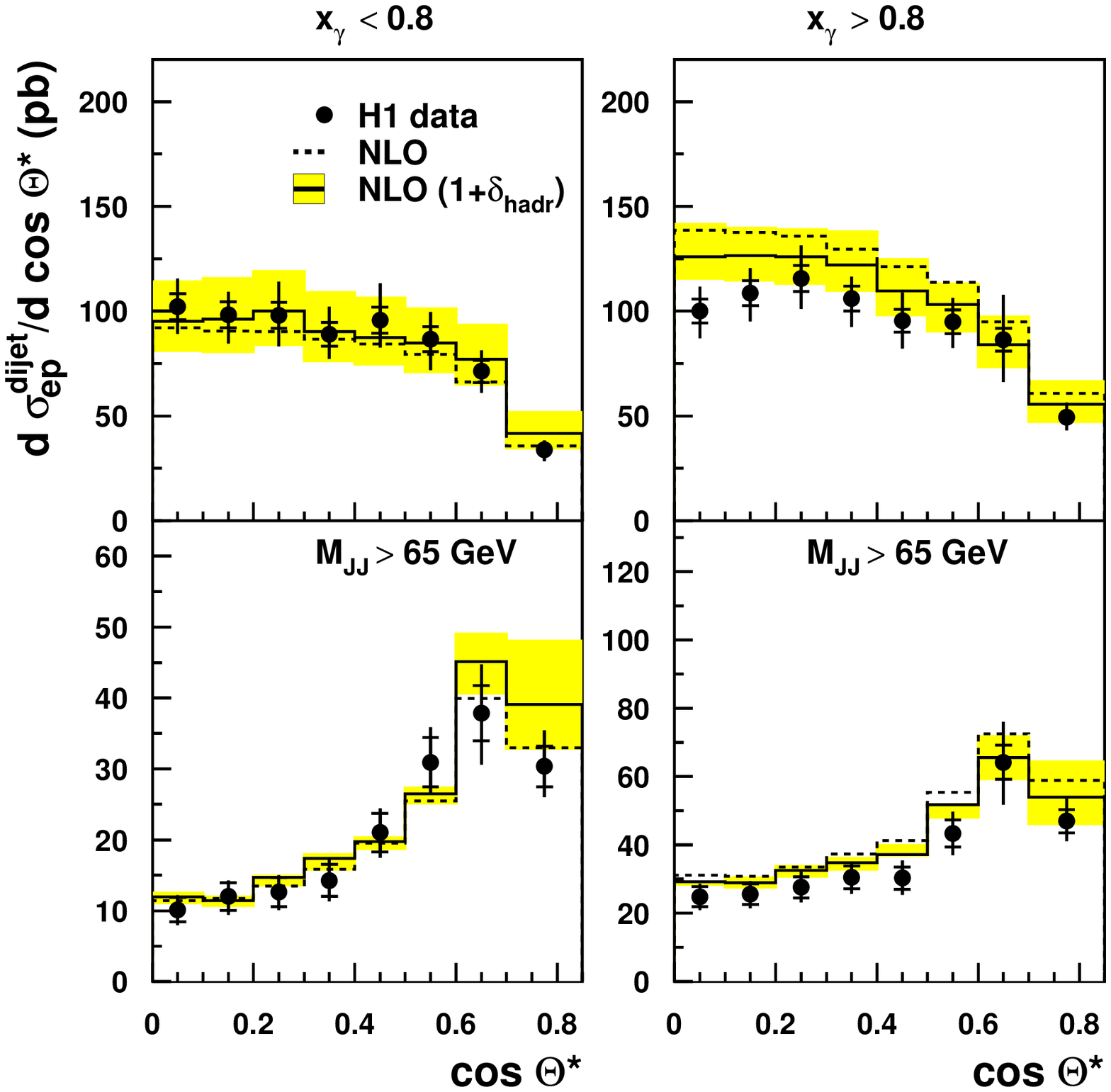,width=\textwidth}
\put(-135,85){\Large{a)}}
\put(-67,85){\Large{b)}}
\put(-135,45){\Large{c)}}
\put(-67,45){\Large{d)}}
\caption{Differential $ep$ cross sections for dijet production 
  ($Q^2< 1\mbox{GeV}^2$) as a 
  function of
  $\cos \theta^*$ distinguished for small $x_{\gamma}$ a) and c) and
  large $x_{\gamma}$ b) and d).
  Figures c) and d) show the cross sections for 
  large invariant masses of the dijet system.
  NLO predictions
  using CTEQ5M pdfs for the proton and
  GRV-HO pdfs for the photon are
  shown as a dashed line.
  The full line shows the NLO predictions, including hadronization corrections  and the grey band
  indicates the renormalization and factorization
  scale uncertainties of the NLO prediction.\label{fig:fig8}}

\end{figure}

Finally, the dijet cross section $d \sigma / d \cos \theta^*$
is plotted in
Figures~\ref{fig:fig8}a) and b) for $x_{\gamma}<0.8$ and $x_{\gamma}>0.8$ respectively.
The cross section decreases with increasing $\cos \theta^*$ 
mainly because of the cuts in $E_T$.
Again, the data are well described by NLO QCD for low $x_{\gamma}$, whereas
at higher $x_{\gamma}$ the predictions overshoot the data for
small values of $\cos \theta^*$.
These cross sections are also shown with a cut on
the invariant mass $M_{JJ}$ of the dijet system in
Figures~\ref{fig:fig8}c) and d),
essentially excluding the
first bin of Figure~\ref{fig:fig1}.
The cut reduces the restriction of the phase space 
due to the correlation with the $E_T$ requirements and changes the shape of the 
distribution towards that expected from the QCD matrix elements.
The QCD calculations
reproduce this transition nicely in both $x_{\gamma}$ regions
where resolved and direct 
photon induced processes contribute with different weights.

\section{Conclusions}
New measurements of dijet cross sections in photoproduction at high
transverse energies are presented for various
jet kinematic observables. The measurements cover invariant dijet masses
up to $180 \gev$ and transverse energies up to $80 \gev$,
reaching $x_p$ and $x_{\gamma}$ values where the experimental
information was previously limited.
In this kinematic domain non-perturbative effects like
multiple interactions and hadronization are found to be
small, which allows a direct comparison
of NLO QCD calculations with the data to be made.
The results demonstrate the power of perturbative
QCD in predicting the measured cross sections
in a wide kinematical range.
Even though the 
photon pdfs have been obtained
from measurements at lower scales,
their QCD evolution correctly reproduces the
data at high scales.
The data do not require significant changes in the parameterizations
of the pdfs but are certainly useful to further constrain the existing ones.
Likewise our understanding of the proton structure in the high $x_p$, high scale region can be improved with the help of these data.
A future stronger constraint 
requires a reduction
of both the theoretical scale uncertainties and the
systematic uncertainties in the data.

\section*{Acknowledgements}

We are grateful to the HERA machine group whose outstanding efforts
have made and continue to make this experiment possible.  We thank the
engineers and technicians for their work in constructing and now
maintaining the H1 detector, our funding agencies for financial
support, the DESY technical staff for continual assistance, and the
DESY directorate for the hospitality which they extend to the non-DESY
members of the collaboration. We wish to thank
S. Frixione and B. P\"{o}tter for many helpful discussions.

\newpage

%%%%%%%%%%%%%%%%%%%%%%%%%%%%%%%%%%%%%%%%%%%%%%%%%%%%%%%%%%%

%%%%%%%%%%%%%%%%%%%%%%%%%%%%%%%%%%%%%%%%%%%%%%%%%%%%%%%%%%%%

%%%%%%%%%%%%%%%%%%%%%%%%%%%%%%%%%%%%%%%%%%%%%%%%%%%%%%%%%%%%

%%%%%%%%% tabellen

\newpage

%\begin{table}[htbp]
%  \begin{center}
%    \begin{tabular}[h]{l}
%
%
%    \end{tabular}
%    \caption{dkdkd}
%    \label{tab:gg}
%  \end{center}
%\end{table}

\begin{table}[htbp]
\begin{center}
\begin{tabular}{|c||c|c|c|c||}

\hline
$M_{JJ}$ (GeV) & $\frac{d\sigma^{\tiny{dijets}}}{dM_{JJ}}$ (pb/GeV)& $\delta_{stat}$(\%) & $+\delta_{tot}$(\%) &$-\delta_{tot}$(\%)\\
\hline
45.-57.5 & 4.30 & 2.5 & 13  & 14\\
57.5-70. & 3.69 & 2.9  & 11 & 11\\
70.-90. & 1.33 & 3.7 & 13 & 11\\
90.-110. & 0.39 & 6.9    &12 & 16\\
110.-135. & 0.101 & 12.  & 19& 18\\
135.-180. & 0.0102 & 27.1   & 32 & 30\\
\hline
$E_{T,mean}$ (GeV) & $\frac{d\sigma^{\tiny{dijets}}}{dE_{T,mean}}$ (pb/GeV)& $\delta_{stat}$(\%) & $+\delta_{tot}$(\%) &$-\delta_{tot}$(\%)\\
\hline
20.-30. &  10.65    &  1.8    &  12    &  12    \\ 
30.-45. &  2.41    &  3.1    &  12    &  11    \\
45.-60. &  0.166    &  11.6    &  16    &  17    \\
60.-80  &  0.0192    &  29.2    &  37    &  34    \\
\hline
$E_{T,max}$ (GeV) & $\frac{d\sigma^{\tiny{dijets}}}{dE_{T,max}}$ (pb/GeV)& $\delta_{stat}$(\%) & $+\delta_{tot}$(\%) &$-\delta_{tot}$(\%)\\
\hline
25.-35. & 12.36 & 1.8  & 11  & 12 \\
35.-45. & 1.82 & 3.9  & 12  & 12  \\
45.-60. & 0.252 & 8.4  & 15  & 15  \\
60.-80. & 0.0198 & 27.4  & 33  & 32 \\
\hline
\hline 
\end{tabular}
\caption{Differential $ep$ cross sections for dijet production as a function 
  of
  the invariant dijet mass $M_{JJ}$ (upper table), as a function of 
  $E_{T,mean}$ (middle table) and as a function of $E_{T,max}$ 
  (lower table) with
  statistical and total upper and lower uncertainties (cf.
  Figures~\ref{fig:fig1} and \ref{fig:fig2}).\label{tab:MJJ}} 
\end{center}
\end{table}

\begin{table}[htbp]
\begin{center}
\begin{tabular}{|c||c|c|c|c||}

\hline
$\overline{\eta}$ & $\frac{d\sigma^{\tiny{dijets}}}{d\overline{\eta}}$ (pb)& $\delta_{stat}$(\%) & $+\delta_{tot}$(\%) &$-\delta_{tot}$(\%)\\
\hline
 & \multicolumn{4}{c||}{$0.1<y<0.5$ and $25<E_{T,max}<35$ GeV} \\
\hline
 0.6-0.9         &  7.55    &  10.0    &  20    &  22    \\
 0.9-1.3         &  42.2    &  4.4    &  11    &  15    \\
 1.3-1.7         &  52.2    &  4.2    &  10    &  11    \\
 1.7-2.1         &  31.8    &  5.4    &  12    &  11    \\
 2.1-2.5         &  10.3    &  10.1    &  17    &  18    \\
\hline
& \multicolumn{4}{c||}{$0.1<y<0.5$ and $35<E_{T,max}<80$ GeV} \\
\hline
 0.9-1.3      &  2.57    &  14.0    &  20    &  20    \\
 1.3-1.7      &  10.2    &  8.0    &  14    &  13    \\
 1.7-2.1      &  6.80    &  9.9    &  14    &  16    \\
 2.1-2.5      &  1.47    &  21.0    &  24    &  24    \\
\hline
& \multicolumn{4}{c||}{$0.5<y<0.9$ and $25<E_{T,max}<35$ GeV} \\
\hline
 0.0-0.6 &  27.9    &  4.7    &  17    &  14    \\
 0.6-0.9 &  66.4    &  4.5    &  13    &  13    \\
 0.9-1.3 &  43.9    &  4.9    &  14    &  11    \\
 1.3-1.7 &  20.8    &  7.4    &  13    &  16    \\
 1.7-2.1 &  10.2    &  11.7    &  14    &  14    \\
 2.1-2.5 &  2.75    &  24.8    &  30    &  29    \\
\hline
& \multicolumn{4}{c||}{$0.5<y<0.9$ and $35<E_{T,max}<80$ GeV} \\
\hline
 0.6-0.9 &  9.77    &  9.7    &  20    &  20    \\
 0.9-1.3 &  14.14    &  7.5    &  13    &  13    \\
 1.3-1.7 &  9.12    &  9.3    &  15    &  15    \\
 1.7-2.1 &  2.59    &  18.1    &  24    &  24   \\
\hline
\hline

\end{tabular}
\caption{Differential $ep$ cross sections for dijet production as a function 
  of $\overline{\eta}$ with
  statistical and total upper and lower uncertainties (cf. 
  Figure~\ref{fig:fig3}).\label{tab:eta}} 
\end{center}
\end{table}

\begin{table}[htbp]
\begin{center}
\begin{tabular}{|c||c|c|c|c||}

\hline
$x_{\gamma}$ & $\frac{d\sigma^{\tiny{dijets}}}{dx_{\gamma}}$ (pb)& $\delta_{stat}$(\%) & $+\delta_{tot}$(\%) &$-\delta_{tot}$(\%)\\
\hline
& \multicolumn{4}{c||}{$x_p<0.1$} \\
\hline
% 0.3-0.5 &  1.682155    &  15.78175    &  17.76918    &  20.07421    \\
 0.5-0.7 &  33.5    &  7.5    &  20    &  20    \\
 0.7-0.85 & 75.0  &  5.8    &  19    &  14    \\
 0.85-1. & 182.7    &  3.3    &  13    &  13    \\
\hline
& \multicolumn{4}{c||}{$x_p>0.1$} \\
\hline  
0.1-0.3  &  41.4   &  8.0    &  15    &  16    \\
 0.3-0.5  &  78.9    &  5.1    &  13    &  13    \\
 0.5-0.7  &  87.2    &  4.5    &  12    &  12    \\
 0.7-0.85 &  126.8  &  4.3    &  12    &  12    \\
 0.85-1   &  256.1   &  3.0    &  10    &  12    \\
\hline
\hline
\end{tabular}
\caption{Differential $ep$ cross sections for dijet production as a function 
  of $x_{\gamma}$ with
  statistical and total upper and lower uncertainties (cf. 
  Figure~\ref{fig:fig4}).\label{tab:xg}} 
\end{center}
\end{table}

\begin{table}[htbp]
\begin{center}
\begin{tabular}{|c||c|c|c|c||}
\hline
$x_p$ & $\frac{d\sigma^{\tiny{dijets}}}{dx_p}$ (pb)& $\delta_{stat}$(\%) & $+\delta_{tot}$(\%) &$-\delta_{tot}$(\%)\\
\hline
& \multicolumn{4}{c||}{$x_{\gamma}<0.8$} \\
\hline
 0.05-0.1 &   288.0  &  5.3    &  15    &  17    \\
 0.1-0.15 &  352.1    &  4.8    &  13    &  11    \\
 0.15-0.22 &  298.6   &  4.4    &  9   &  15    \\
 0.22-0.32 &  121.8  &  5.6   &  19    &  12    \\
 0.32-0.6 &  7.86  &  11.8    &  17    &  19    \\
\hline
& \multicolumn{4}{c||}{$x_{\gamma}>0.8$} \\
\hline
 0.05-0.1 &  530.0    &  3.5    &  12    &  14    \\
 0.1-0.15 &  384.4  &  4.3    &  12    &  14    \\
 0.15-0.22 &  232.1    &  4.7    &  11    &  12    \\
 0.22-0.32 &    83.7   &  6.5    &  13    &  13    \\
 0.32-0.6 &  7.4    &  12.7    &  19    &  18    \\
\hline
\hline
\end{tabular}
\caption{Differential $ep$ cross sections for dijet production as a function 
  of $x_p$ with
  statistical and total upper and lower uncertainties (cf. 
  Figure~\ref{fig:fig4}).\label{tab:xp}} 
\end{center}
\end{table}

\begin{table}[htbp]
\begin{center}
\begin{tabular}{|c||c|c|c|c||}
\hline
$x_{\gamma}$ & $\frac{d\sigma^{\tiny{dijets}}}{dx_{\gamma}}$ (pb)& $\delta_{stat}$(\%) & $+\delta_{tot}$(\%) &$-\delta_{tot}$(\%)\\
\hline
 & \multicolumn{4}{c||}{$25<E_{T,max}<35$ GeV} \\
\hline
 0.1-0.3 &  40.1    &  8.4    &  17    &  18   \\
 0.3-0.5 &  77.3    &  5.3    &  11    &  13    \\
 0.5-0.7 &  101.1    &  4.3    &  14    &  14    \\
 0.7-0.85 & 173.6    &  3.9   &  14    &  11    \\
 0.85-1. &  361.7    &  2.5    &  11    &  12    \\
\hline
 & \multicolumn{4}{c||}{$35<E_{T,max}<80$ GeV} \\
\hline
 0.1-0.3 & 2.34    &  27.5    &  34    &  31    \\
 0.3-0.5 &  9.83    &  12.1    &  17    &  18    \\
 0.5-0.7 &  19.7    &  8.5    &  14    &  15    \\
 0.7-0.85 &  28.7    &  8.0    &  15    &  16    \\
 0.85-1.  &  78.1    &  4.9    &  13    &  12    \\
\hline
\hline
\end{tabular}
\caption{Differential $ep$ cross sections for dijet production as a function 
  of $x_{\gamma}$ with
  statistical and total upper and lower uncertainties (cf. 
  Figures~\ref{fig:fig5}, \ref{fig:fig6} and \ref{fig:fig7}).\label{tab:xgET}} 
\end{center}
\end{table}

\begin{table}[htbp]
\begin{center}
\begin{tabular}{|c||c|c|c|c||}
\hline
$\cos \theta^*$ & $\frac{d\sigma^{\tiny{dijets}}}{d\cos \theta^*}$ (pb)& $\delta_{stat}$(\%) & $+\delta_{tot}$(\%) &$-\delta_{tot}$(\%)\\
\hline
& \multicolumn{4}{c||}{$x_{\gamma}<0.8$} \\
\hline
 0.0-0.1 &   102.0   &  6.1    &  13    &  13    \\
 0.1-0.2 &  98.3    &  6.3    &  11    &  14    \\
 0.2-0.3 &  98.0   &  6.4    &  16    &  15    \\
 0.3-0.4 &  89.0    &  6.4   &  15    &  13    \\
 0.4-0.5 &  95.6    &  6.5    &  18    &  14    \\
 0.5-0.6 &  86.6    &  6.8    &  15    &  17    \\
 0.6-0.7 &  71.3    &  7.5    &  14    &  14    \\
 0.7-0.85 &  33.8    &  8.8    &  14    &  16    \\
\hline
& \multicolumn{4}{c||}{$x_{\gamma}>0.8$} \\
\hline 
 0.0-0.1 &  100.1    &  5.6    &  12    &  13    \\
 0.1-0.2 &  108.6    &  5.5    &  11    &  12    \\
 0.2-0.3 &  115.6    &  5.4    &  14    &  13    \\
 0.3-0.4 &  106.1    &  5.5    &  10    &  13    \\
 0.4-0.5 &  95.7   &  5.7    &  14    &  14    \\
 0.5-0.6 &  95.0    &  5.9    &  12    &  13    \\
 0.6-0.7 &  86.3    &  6.3    &  25    &  23    \\
 0.7-0.85 &  49.5    & 7.0    &  14    &  13    \\
\hline
& \multicolumn{4}{c||}{$x_{\gamma}<0.8$ and $M_{JJ}>65$ GeV} \\ 
\hline
 0.0-0.1 &  10.1    &  17.0    &  21    &  21    \\
 0.1-0.2 &  12.0    &  16.4    &  19    &  21    \\
 0.2-0.3 &  12.6  &  16.3    &  19    &  20    \\
 0.3-0.4 &  14.3    &  15.9    &  22    &  21    \\
 0.4-0.5 &  21.0    &  12.8    &  16    &  17    \\
 0.5-0.6 &  30.9    &  11.3    &  16    &  14    \\
 0.6-0.7 &  37.9    &  10.3    &  18    &  19    \\
 0.7-0.85 & 30.4    &  9.4    &  17    &  14    \\
\hline
& \multicolumn{4}{c||}{$x_{\gamma}>0.8$ and $M_{JJ}>65$ GeV} \\ 
\hline
0.0-0.1 &  24.9   &  11.7    &  15    &  16    \\
0.1-0.2 &  25.6    &  11.9    &  15    &  16    \\
0.2-0.3 &  27.6    &  11.3    &  16    &  16    \\
0.3-0.4 &  30.5    &  11.0    &  15    &  16    \\
0.4-0.5 &  30.3    &  10.9    &  17    &  16    \\
0.5-0.6 &  43.4    &  9.2    &  15    &  15    \\
0.6-0.7 &  64.2    &  7.7    &  19    &  19    \\
0.7-0.85 &  47.0    &  7.4    &  14    &  13    \\
\hline
\hline 
\end{tabular}
\caption{Differential $ep$ cross sections for dijet production as a function 
  of $\cos \theta^*$ with
  statistical and total upper and lower uncertainties (cf. 
  Figure~\ref{fig:fig8}).\label{tab:costh}} 
\end{center}
\end{table}

%%%%%%%%%%%%%%%%%%%%%%%%%%%%%%%%%%%%%%%%%%%%%%%%%%%%%%%%%%%


\begin{thebibliography}{99}


%\cite{Ahmed:1992xj}
\bibitem{Ahmed:1992xj}
T.~Ahmed {\it et al.}  [H1 Collaboration],
%``Hard scattering in gamma p interactions,''
Phys.\ Lett.\ B {\bf 297} (1992) 205.
%%CITATION = PHLTA,B297,205;%%

%\cite{Adloff:1998da}
\bibitem{Adloff:1998da}
C.~Adloff {\it et al.}  [H1 Collaboration],
%``Measurement of the inclusive di-jet cross section in photoproduction  and determination of an effective parton distribution in the photon,''
Eur.\ Phys.\ J.\ C {\bf 1} (1998) 97
[arXiv:hep-ex/9709004].
%%CITATION = HEP-EX 9709004;%%

%\cite{Adloff:2000bs}
\bibitem{Adloff:2000bs}
C.~Adloff {\it et al.}  [H1 Collaboration],
%``Measurement of di-jet cross-sections in photoproduction and photon  structure,''
Phys.\ Lett.\ B {\bf 483} (2000) 36
[arXiv:hep-ex/0003011].
%%CITATION = HEP-EX 0003011;%%


%\cite{Breitweg:1998rx}
\bibitem{Breitweg:1998rx}
J.~Breitweg {\it et al.}  [ZEUS Collaboration],
%``Dijet cross sections in photoproduction at HERA,''
Eur.\ Phys.\ J.\ C {\bf 1} (1998) 109
[arXiv:hep-ex/9710018].
%%CITATION = HEP-EX 9710018;%%


%\cite{Breitweg:1999wi}
\bibitem{Breitweg:1999wi}
J.~Breitweg {\it et al.}  [ZEUS Collaboration],
%``Measurement of dijet photoproduction at high transverse energies at  HERA,''
Eur.\ Phys.\ J.\ C {\bf 11} (1999) 35
[arXiv:hep-ex/9905046].
%%CITATION = HEP-EX 9905046;%%


%\cite{Affolder:2001ew}
\bibitem{Affolder:2001ew}
T.~Affolder {\it et al.}  [CDF Collaboration],
%``Measurement of the two-jet differential cross section in proton  antiproton collisions at s**(1/2) = 1800-GeV,''
Phys.\ Rev.\ D {\bf 64} (2001) 012001
[arXiv:hep-ex/0012013].
%%CITATION = HEP-EX 0012013;%%

%\cite{Abazov:2001hb}
\bibitem{Abazov:2001hb}
V.~M.~Abazov {\it et al.}  [D0 Collaboration],
%``The inclusive jet cross-section in p anti-p collisions at s**(1/2) =  1.8-TeV using the k(T) algorithm,''
arXiv:hep-ex/0109041.
%%CITATION = HEP-EX 0109041;%%

%\cite{Nisius:2000cv}
\bibitem{Nisius:2000cv}
R.~Nisius,
%``The photon structure from deep inelastic electron photon scattering,''
Phys.\ Rept.\  {\bf 332} (2000) 165
[arXiv:hep-ex/9912049].
%%CITATION = HEP-EX 9912049;%%

%\cite{Huston:1998jj}
\bibitem{Huston:1998jj}
J.~Huston, S.~Kuhlmann, H.~L.~Lai, F.~I.~Olness, J.~F.~Owens, D.~E.~Soper and W.~K.~Tung,
%``Study of the uncertainty of the gluon distribution,''
Phys.\ Rev.\ D {\bf 58} (1998) 114034
[arXiv:hep-ph/9801444].
%%CITATION = HEP-PH 9801444;%%

\bibitem{ZEUSDiJet}
S.~Chekanov {\it et al.}  [ZEUS Collaboration],
DESY~01-220, arXiv:hep-ex/0112029.

\bibitem{wwa}
P. Kessler, Il Nuovo Cimento {\bf 17} (1960) 809.

%\cite{Budnev:1974de}
\bibitem{Budnev:1974de}
V.~M.~Budnev, I.~F.~Ginzburg, G.~V.~Meledin and V.~G.~Serbo,
%``The Two Photon Particle Production Mechanism. Physical Problems. Applications. Equivalent Photon Approximation,''
Phys.\ Rept.\  {\bf 15} (1974) 181.
%%CITATION = PRPLC,15,181;%%

%\cite{Frixione:1993yw}
\bibitem{Frixione:1993yw}
S.~Frixione, M.~L.~Mangano, P.~Nason and G.~Ridolfi,
%``Improving the Weizsacker-Williams approximation in electron - proton collisions,''
Phys.\ Lett.\ B {\bf 319} (1993) 339
[arXiv:hep-ph/9310350].
%%CITATION = HEP-PH 9310350;%%

%\cite{Frixione:1997ks}
\bibitem{Frixione:1997ks}
S.~Frixione and G.~Ridolfi,
%``Jet photoproduction at HERA,''
Nucl.\ Phys.\ B {\bf 507} (1997) 315
[arXiv:hep-ph/9707345].
%%CITATION = HEP-PH 9707345;%%

%\cite{Klasen:1997it}
\bibitem{Klasen:1997it}
M.~Klasen and G.~Kramer,
%``Inclusive two-jet production at HERA: Direct and resolved  cross sections in next-to-leading order QCD,''
Z.\ Phys.\ C {\bf 76} (1997) 67
[arXiv:hep-ph/9611450].
%%CITATION = HEP-PH 9611450;%%

%\cite{Harris:1997hz}
\bibitem{Harris:1997hz}
B.~W.~Harris and J.~F.~Owens,
%``Photoproduction of jets at HERA in next-to-leading order QCD,''
Phys.\ Rev.\ D {\bf 56} (1997) 4007
[arXiv:hep-ph/9704324].
%%CITATION = HEP-PH 9704324;%%

%\cite{Aurenche:2000nc}
\bibitem{Aurenche:2000nc}
P.~Aurenche, L.~Bourhis, M.~Fontannaz and J.~P.~Guillet,
%``NLO Monte Carlo approach in 1 or 2 jets photoproduction,''
Eur.\ Phys.\ J.\ C {\bf 17} (2000) 413
[arXiv:hep-ph/0006011].
%%CITATION = HEP-PH 0006011;%%

%\cite{Baur:1992pp}
\bibitem{Baur:1992pp}
U.~Baur, J.~A.~Vermaseren and D.~Zeppenfeld,
%``Electroweak vector boson production in high-energy e p collisions,''
Nucl.\ Phys.\ B {\bf 375} (1992) 3.
%%CITATION = NUPHA,B375,3;%%

%\cite{Ellis:1993tq}
\bibitem{Ellis:1993tq}
S.~D.~Ellis and D.~E.~Soper,
%``Successive combination jet algorithm for hadron collisions,''
Phys.\ Rev.\ D {\bf 48} (1993) 3160
[arXiv:hep-ph/9305266].
%%CITATION = HEP-PH 9305266;%%

%\cite{Catani:1993hr}
\bibitem{Catani:1993hr}
S.~Catani, Y.~L.~Dokshitzer, M.~H.~Seymour and B.~R.~Webber,
%``Longitudinally invariant K(t) clustering algorithms for hadron-hadron collisions,''
Nucl.\ Phys.\ B {\bf 406} (1993) 187.
%%CITATION = NUPHA,B406,187;%%

%\cite{Adloff:2001tq}
\bibitem{Adloff:2001tq}
C.~Adloff {\it et al.}  [H1 Collaboration],
%``Measurement and QCD analysis of jet cross sections in deep-inelastic  positron proton collisions at s**(1/2) of 300-GeV,''
Eur.\ Phys.\ J.\ C {\bf 19} (2001) 289
[arXiv:hep-ex/0010054].
%%CITATION = HEP-EX 0010054;%%

%\cite{Sjostrand:1994yb}
\bibitem{Sjostrand:1994yb}
T.~Sj\"ostrand,
%``High-energy physics event generation with PYTHIA 5.7 and JETSET 7.4,''
Comput.\ Phys.\ Commun.\  {\bf 82} (1994) 74.
%%CITATION = CPHCB,82,74;%%

%\cite{Marchesini:1992ch}
\bibitem{Marchesini:1992ch}
G.~Marchesini, B.~R.~Webber, G.~Abbiendi, I.~G.~Knowles, M.~H.~Seymour and L.~Stanco,
%``HERWIG: A Monte Carlo event generator for simulating hadron emission reactions with interfering gluons. Version 5.1 - April 1991,''
Comput.\ Phys.\ Commun.\  {\bf 67} (1992) 465.
%%CITATION = CPHCB,67,465;%%

%\cite{Brun:1987ma}
\bibitem{Brun:1987ma}
R.~Brun, F.~Bruyant, M.~Maire, A.~C.~McPherson and P.~Zanarini,
%``Geant3,''
CERN-DD/EE/84-1.

%\cite{Gluck:1994uf}
\bibitem{Gluck:1994uf}
M.~Gl\"uck, E.~Reya and A.~Vogt,
%``Dynamical parton distributions of the proton and small x physics,''
Z.\ Phys.\ C {\bf 67} (1995) 433.
%%CITATION = ZEPYA,C67,433;%%

%\cite{Gluck:1992jc}
\bibitem{Gluck:1992jc}
M.~Gl\"uck, E.~Reya and A.~Vogt,
%``Photonic parton distributions,''
Phys.\ Rev.\ D {\bf 46} (1992) 1973.
%%CITATION = PHRVA,D46,1973;%%

%\cite{Frixione:1997np}
\bibitem{Frixione:1997np}
S.~Frixione,
%``A general approach to jet cross sections in QCD,''
Nucl.\ Phys.\ B {\bf 507} (1997) 295
[arXiv:hep-ph/9706545].
%%CITATION = HEP-PH 9706545;%%

%\cite{Lai:1997mg}
\bibitem{Lai:1997mg}
H.~L.~Lai {\it et al.},
%``Improved parton distributions from global analysis of recent deep  inelastic scattering and inclusive jet data,''
Phys.\ Rev.\ D {\bf 55} (1997) 1280
[arXiv:hep-ph/9606399].
%%CITATION = HEP-PH 9606399;%%

%\cite{Martin:2000ww}
\bibitem{Martin:2000ww}
A.~D.~Martin, R.~G.~Roberts, W.~J.~Stirling and R.~S.~Thorne,
%``Parton distributions and the LHC: W and Z production,''
Eur.\ Phys.\ J.\ C {\bf 14} (2000) 133
[arXiv:hep-ph/9907231].
%%CITATION = HEP-PH 9907231;%%

%\cite{Gluck:1992ee}
\bibitem{Gluck:1992ee}
M.~Gl\"uck, E.~Reya and A.~Vogt,
%``Parton structure of the photon beyond the leading order,''
Phys.\ Rev.\ D {\bf 45} (1992) 3986.
%%CITATION = PHRVA,D45,3986;%%

%\cite{Aurenche:1994in}
\bibitem{Aurenche:1994in}
P.~Aurenche, J.~P.~Guillet and M.~Fontannaz,
%``Parton distributions in the photon,''
Z.\ Phys.\ C {\bf 64} (1994) 621
[arXiv:hep-ph/9406382].
%%CITATION = HEP-PH 9406382;%%

%\cite{Abt:hi}
\bibitem{Abt:hi}
I.~Abt {\it et al.}  [H1 Collaboration],
%``The H1 Detector At Hera,''
Nucl.\ Instrum.\ Meth.\ A {\bf 386} (1997) 310.
%%CITATION = NUIMA,A386,310;%%

%\cite{Abt:1996xv}
\bibitem{Abt:1996xv}
I.~Abt {\it et al.}  [H1 Collaboration],
%``The Tracking, calorimeter and muon detectors of the H1 experiment at HERA ,''
Nucl.\ Instrum.\ Meth.\ A {\bf 386} (1997) 348.
%%CITATION = NUIMA,A386,348;%%

%\cite{Andrieu:1993kh}
\bibitem{Andrieu:1993kh}
B.~Andrieu {\it et al.}  [H1 Calorimeter Group Collaboration],
%``The H1 liquid argon calorimeter system,''
Nucl.\ Instrum.\ Meth.\ A {\bf 336} (1993) 460.
%%CITATION = NUIMA,A336,460;%%


%\cite{Appuhn:1996na}
\bibitem{Appuhn:1996na}
R.~D.~Appuhn {\it et al.}  [H1 SPACAL Group Collaboration],
%``The H1 lead/scintillating-fibre calorimeter,''
Nucl.\ Instrum.\ Meth.\ A {\bf 386} (1997) 397.
%%CITATION = NUIMA,A386,397;%%

%\cite{Adloff:1997mi}
\bibitem{Adloff:1997mi}
C.~Adloff {\it et al.}  [H1 Collaboration],
%``Diffraction dissociation in photoproduction at HERA,''
Z.\ Phys.\ C {\bf 74} (1997) 221
[arXiv:hep-ex/9702003].
%%CITATION = HEP-EX 9702003;%%

\bibitem{yJB} A. Blondel and F. Jacquet, Proceedings of the Study of
  an $ep$
  Facility for Europe\\
  ed. U. Amaldi, DESY 79/48 (1979) 391.


%\cite{Lonnblad:1992tz}
\bibitem{Lonnblad:1992tz}
L.~L\"onnblad,
%``ARIADNE version 4: A Program for simulation of QCD cascades implementing the color dipole model,''
Comput.\ Phys.\ Commun.\  {\bf 71} (1992) 15.
%%CITATION = CPHCB,71,15;%%



%\cite{Charchula:1994kf}
\bibitem{Charchula:1994kf}
K.~Charchula, G.~A.~Schuler and H.~Spiesberger,
%``Combined QED and QCD radiative effects in deep inelastic lepton - proton scattering: The Monte Carlo generator DJANGO6,''
Comput.\ Phys.\ Commun.\  {\bf 81} (1994) 381.
%%CITATION = CPHCB,81,381;%%

\end{thebibliography}
\end{document}